\documentclass[useAMS,usenatbib]{mn2e}
\usepackage{threeparttable}
\usepackage{amsmath}
\usepackage{amssymb}
\usepackage{url}
\usepackage{graphicx}
\newcommand{\be}{\begin{equation}}
\newcommand{\ee}{\end{equation}}
\newcommand{\bea}{\begin{eqnarray}}
\newcommand{\eea}{\end{eqnarray}}

\def\apj{ApJ}

\def\arcsec{^{\prime\prime}}

\usepackage{color}              
\usepackage{epsfig}
\usepackage{amsmath}
\usepackage{amssymb}
\usepackage{amstext}

\graphicspath{
{figures/}
}
\title[Bullet Cluster: Remarkable Revelations and Enduring Insights -- Part I]{Bullet Cluster: Remarkable Revelations and Enduring Insights -- Part I}
\author[Siddharth Savyasachi Malu and Sanjay K. Pandey]{Siddharth Savyasachi Malu$^{1,3}$\thanks{E-mail: siddharth@iiti.ac.in}
  and Sanjay K. Pandey$^{2,4}$\\
$^{1}$Dept. of Astronomy, Astrophysics and Space Engineering, IIT Indore, Simrol, Khandwa Road, Indore 453552 India\\
$^{2}$Sri Lal Bahadur Shastri Degree College, Gonda India}
\begin{document}
\maketitle
\label{firstpage}
\begin{abstract}
This study unveils new insights into the Bullet Cluster through cm-wave observations made with the Australia Telescope Compact Array, focusing on the Sunyaev-Zel’dovich Effect (SZE). A novel SZE feature was discovered, distinctively offset from the X-ray brightness peak, challenging conventional associations between such phenomena. The findings, achieved through high-resolution interferometric imaging, reveal complex substructures within the SZE, including significantly displaced components uncorrelated with X-ray, optical, or lensing maps. This deviation underscores the complexity within the cluster's dynamics and suggests that detailed, high-resolution SZE imaging is crucial for understanding the physics of merging clusters. We perform several tests to ensure that one of these novel and unusual features is indeed real, and that the detection of all features in the images is robust. In addition to these displacements, we find several SZE features that are tentatively located at critical boundaries in this merging cluster - these features will be discussed in the next few manuscripts in this series. These observations pave the way for more sophisticated analyses, combining SZE and X-ray data, to decode the evolutionary mysteries of cosmic structures like the Bullet Cluster. We note here that we are discussing only one novel / peculiar feature in the observations of the cluster done in the year 2019. Further observations have revealed more features, which we shall discuss in the next few manuscripts in this series on the Bullet cluster.
\end{abstract}
\begin{keywords}
cosmic background radiation --- galaxies: clusters: individual (1E
0657--56, RX J0658--5557) ---intergalactic medium --- radio continuum:
general --- techniques: interferometric
\end{keywords}
\section{Introduction}
\label{intro}
The Sunyaev--Zeldovich effect (SZE) is the inverse Compton scattering of CMB
photons off a collection of electrons which causes a spectral
distortion in the CMB \citep{1972CoASP...4..173S}. This inverse Compton
scattering causes a decrease in CMB brightness at frequencies below
217 GHz and an increase above 217 GHz. 
The amplitude of the decrease
in brightness is given by $\Delta T\propto \int
n_{e}T_{e}d\ell$\footnote{$T_e=$Electron temperature; $n_e=$electron
  number density}, where the integral is along the line of sight. Galaxy clusters are the
largest gravitationally bound objects and the largest sources
of hot, dense ionized gas, which makes them ideal targets for studying
through the SZE, especially since the SZE is independent of redshift.

1E0657$-$56, known as the {\bf Bullet Cluster}, is one of the hottest
known clusters that has been well-studied over the last decade for a
variety of reasons; namely, the existence of a strong radio halo \citep{2000ApJ...544..686L}, the Sunyaev-Zel'dovich 
effect (\citet{2009ApJ...701...42H}, \citet{2010ApJ...716.1118P} and references therein), its X-ray brightness \citep{2002ApJ...567L..27M}, though most notably in providing the
most direct proof of the existence of dark matter \citep{2006ApJ...648L.109C}. It is
a cluster collision/merger event at $z\sim$0.296, with the larger,
westward cluster being $\sim$10 times the mass of the smaller `bullet'.

Cluster mergers have by now been observed in many different regimes:
from X--ray observations using the {\em Chandra} to radio
observations. Diffuse non-thermal radio emission in clusters of galaxies of
$\sim$Mpc size not associated with galaxies, when close to the centres of
clusters, are called {\em radio halos}, and when at or close to the
peripheries of clusters are called {\em radio relics}. These are associated with 
relativistic electrons and magnetic fields in the ICM and have steep
synchrotron spectra ($S\propto\nu^{-\alpha}$,$\alpha >$1.0); radio
relics have steeper spectra compared to radio halos. About 35\% of
clusters with X-ray luminosities greater than $5\times10^{44}$erg s$^{-1}$
have radio halos (Venturi et al 2008). However, among clusters that
are dynamically disturbed, about 75\% exhibit radio halos (Cassano
et al 2010). The total power radiated from radio halos (at 1.4 GHz) is
found to scale with X-ray luminosity and temperature. In the currently favoured model of radio halo formation, the `turbulent re--acceleration through merger' model, the spectra of radio halos is expected and indeed, observed to steepen considerably at frequencies below 10 GHz.  Measurements by Liang et al (2000) suggest such a steepening at $\sim$8.5 GHz. The radio halo in the Bullet cluster was thus not expected to be observable at 18 GHz.

Since not many
clusters at redshifts greater than 0.3 were known prior to the
PLANCK satellite and MACSurvey, it has been difficult to probe the existence of a redshift
evolution in the occurrence of radio halos, until now.

Two competing models exist as explanations of the origins or radio
halos: the so--called {\em Primary} or {\em Turbulent Reacceleration}
(see, e.g. \citet{2011MNRAS.412..817B,2010A&A...517A..10C,2010ApJ...721L..82C})
and {\em Secondary} models (e.g. Keshet et al. 2010). Primary models
of radio halo formation feature an exponential  steepening of the spectrum, or a
``knee'' -- the exact frequency at which the ``knee'' occurs depends
on the epoch of the cluster merger, and the age of the radio
halo. This ``knee'' is seen at a wide range of frequencies, from 1.4
GHz to 5 GHz. In secondary models, the steepening of the spectrum is followed by a
flattening, usually at frequencies higher than 9 GHz. Deep high--frequency observations (above 9 GHz) of radio halos
are therefore crucial for understanding the origins of radio halos. 

In this paper, we present the first 18 GHz interferometer observations of the Bullet
cluster from the Australia Telescope Compact Array (ATCA) using the
recently installed Compact Array Broadband Backend or CABB \citep{2011MNRAS.416..832W}. We
present our observations in \S2, describe the imaging of the cluster at 18 GHz in
\S3, enumerate the consistency checks run by us in \S4, place limits on the SZE
in \S5 and discuss the SZE features in \S6. It may be noted here that the interferometer imaging lacks the low spatial frequencies that are required to image the SZE effect in its entirety, and misses the extended SZE emission; therefore our image may not be useful for fitting gas models along with X-ray data.  The imaging reveals, for the first time, the existence of a SZE component in the Bullet Cluster displaced from the X--ray peak, pointing to the importance and need for high angular resolution imaging of SZE in such complex merging clusters in order to usefully model the cluster gas.

Observations from the South Pole Telescope \citep{2010ApJ...719.1045L} show that on arcminute and smaller scales
observed CMB anisotropy power is significantly less than levels expected from models of SZE from 
a cosmological distribution of clusters. This discrepancy may be an indication of our lack of understanding of
the environmental impact of cluster mergers. This paper presents sub-arcmin angular resolution 
observations of the SZE in the cluster, which demonstrates that the relation between SZE and X-ray emission
is complex and requires additional physics that may be related to the impact
of the merger on intracluster gas.

\section{Radio Observations and Imaging}
\label{radio_obs}
The Australia Telescope Compact Array (ATCA) is a radio interferometer
with six 22-m antennas, five of which may be positioned on stations
along a `T'-shaped rail track that is 3-km along E-W and 214-m along
N-S. 

Observations were made in a pair of 2-GHz bands: a `17-GHz band' covering frequencies 16-18 GHz
and a `19-GHz band' covering the range 18-20 GHz.
Each of the 2-GHz wide bands were subdivided into 2048 frequency
channels. All observations were in full polarization mode and recorded
multi-channel continuum visibilities. 

X-ray images of the cluster show two peaks separated by $\sim 1\farcm5$ and the detected
extent of X-ray emission is $\sim 6\arcmin$ \citep{1999ApJ...513...23A} suggesting that hot
gas pervades a significant area between and beyond the constituent clusters.
Since the FWHM field-of-view of the ATCA antennas is
$2\farcm6$ at 18 GHz, we mosaic imaged the Bullet cluster in 2009
using 2 pointing centres at (J2000 epoch coordinates) RA: $06^{\rm h}58^{\rm m}20^{\rm s}$, 
DEC: $-55\degr56\arcmin28\arcsec$ 
and RA: $06^{\rm h}58^{\rm m}30^{\rm s}$, 
DEC: $-55\degr56\arcmin28\arcsec$ and in 2010 using 4 pointing centres
at  RA: $06^{\rm h}58^{\rm m}32.19^{\rm s}$, 
DEC: $-55\degr57\arcmin00\arcsec$; RA: $06^{\rm h}58^{\rm m}30^{\rm s}$, 
DEC: $-55\degr56\arcmin00\arcsec$;  RA: $06^{\rm h}58^{\rm m}25^{\rm s}$, 
DEC: $-55\degr57\arcmin00\arcsec$ and  RA: $06^{\rm h}58^{\rm m}22.81^{\rm s}$, 
DEC: $-55\degr56\arcmin00\arcsec$. 
%
%

Observations were made in each of two ATCA array configurations: H168
and H75 that have baselines up to 168 and 75 m respectively; the former to
enable subtraction of unresolved continuum sources, the latter to provide
enhanced surface brightness sensitivity for imaging of the
SZE. A journal of the observations is in Table~\ref{obs_journal}. In each of the observing sessions, antenna pointing
corrections were updated every hour using a 5-point offset pattern observation
on PKS B0537$-$441, the unresolved calibrator PKS B0742$-$56 was observed
every 10 min to monitor and correct for amplitude and phase drifts in the
interferometer arms, and PKS B1934$-$638 was observed once every session as a
primary calibrator to set the absolute flux density scale. During the 10 mins
between successive calibrations, the two pointing positions were sequentially
observed for 90 sec each.  Visibilities were recorded with 10 sec averaging.
\begin{center}
\begin{table}
\caption{Journal of the ATCA observations.}
\label{obs_journal}
\begin{tabular}{@{}lcr}
\hline
Array configuration & Observing time (hours) & Date \\
\hline
H168 & 12 & 2009 April 25 \\
H168 & 12 & 2009 April 26 \\
H168 & 12 & 2010 July 29 \\
H168 & 12 & 2010 July 30 \\
H75 & 8 & 2009 June 28 \\
H75 & 8 & 2009 June 29 \\
H75 & 8 & 2009 June 30 \\
H75 & 6 & 2010 September 21 \\
H75 & 6 & 2010 September 22 \\
\hline
\end{tabular}
\end{table}
\end{center}
Interferometer visibilities were examined, calibrated and imaged using MIRIAD; all
image processing were also accomplished using utilities in this software package.
The visibility data in the 17 and 19 GHz bands were separately edited for
interference and calibrated before bandwidth synthesis imaging.
Visibility data in each of the 2 GHz wide bands were recorded over 2048 frequency channels, and 50 channels at
each of the band edges were excluded from analysis to avoid data in frequency
domains where signal path gains are relatively low. Frequency channels that
appeared to have relatively large fluctuations in visibility amplitude owing to hardware 
faults in the digital correlator were also rejected prior to calibration and imaging.
\subsection{Calibration}
\label{gain_leakage}
Adopted fluxes for the primary calibrator PKS~B1934$-$638 were 1.146
and 0.992 Jy in the 17 and 19 GHz bands respectively; the spectral index
$\alpha$ (defined as $S_{\nu} \propto \nu^{\alpha}$, where $S_{\nu}$ is the
flux density at frequency $\nu$)  was adopted to be $-1.33$ in both
bands (see \citet{atca_sault2003}). Outliers in the amplitudes of visibility data on PKS~B1934$-$638 were
rejected---removing 15\% of data---and the reliable visibilities
were used to set the absolute flux density
scale as well as determine the instrument bandpass calibration.
When calibrated for the bandpass, the visibility amplitudes of PKS~B0742$-$56
showed continuity across the 4-GHz observing frequency range and a trend
consistent with a single power-law: this was a check of the bandpass
calibration.  Drifts of up to $30\degr$ were observed in the interferometer arms
over the observing sessions: calibrations for the time-varying complex gains in the antenna signal paths
as well as calibrations for polarization leakages were derived from the visibilities on
PKS~B0742$-$56. RMS phase variations in antenna signal paths within the 1-min
calibrator scans was within $10\degr$, indicating that short timescale atmospheric and
instrumental phase cycling would not result in amplitude attenuation of more
than 3\%. 

Since no circular polarization is expected, Stokes V is expected to be consistent with
thermal noise. Therefore, at times and frequency channels where Stokes V
visibilities deviate more than four times rms thermal noise in the calibrated
visibilities acquired towards the Bullet cluster pointings, data in all
Stokes parameters were rejected. Stokes--V based clipping was therefore done, aimed at automated
rejection of self-generated low level interference.
\begin{figure}
  \begin{center}
    \includegraphics[height=70mm, angle=0]{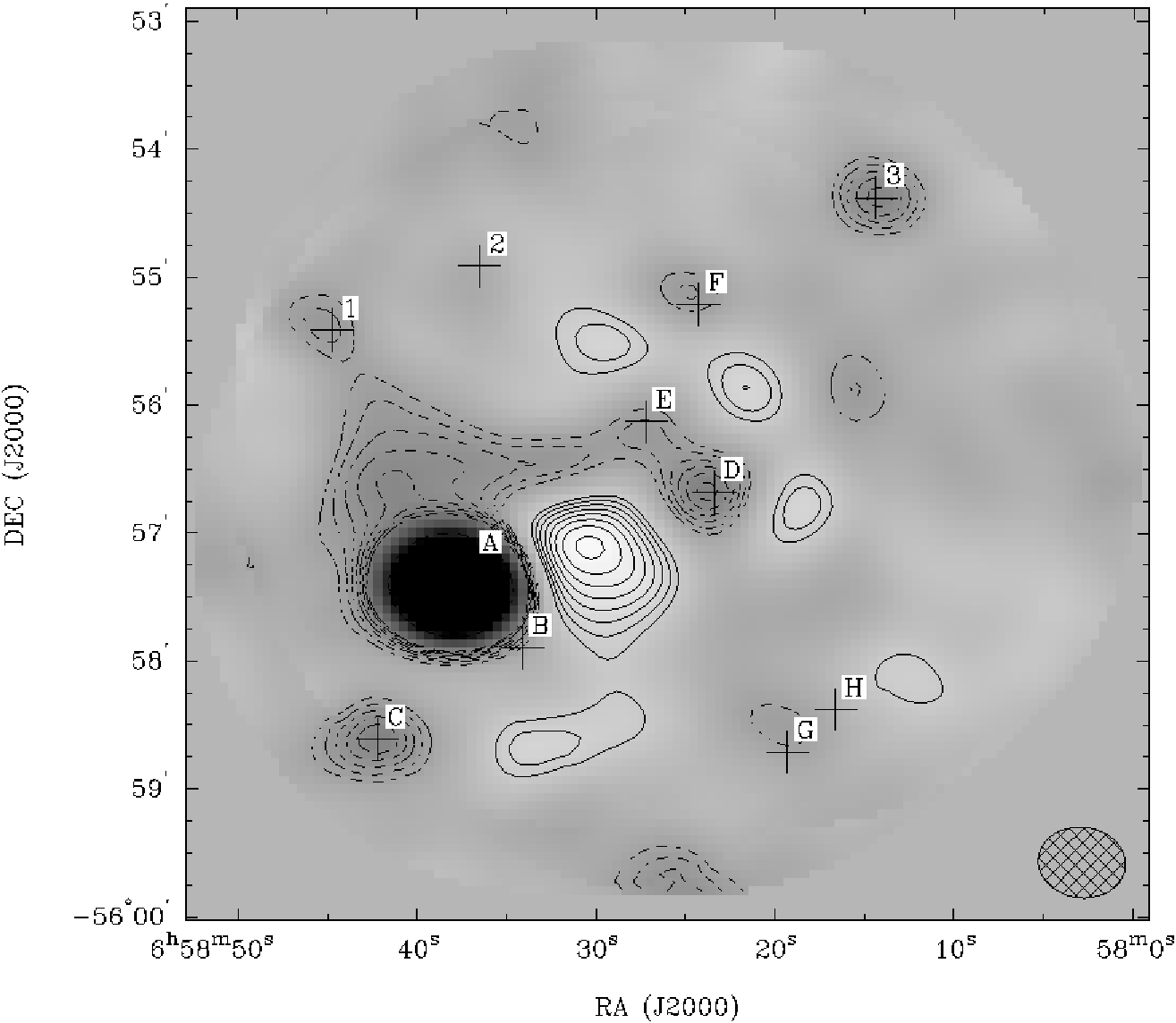}\\
    \caption
        {
          18 GHz mosaic image of the Bullet cluster. In this representation as
          well as subsequent images displayed herein the image has been tapered to make the rms noise uniform over the field
          of view; this results in an attenuation of unresolved sources by factor
          0.02 at the edge. Image rms noise is 6.5~$\mu$Jy~beam$^{-1}$; the beam
          FWHM is $39\farcs8 \times 32\farcs1$ at $87.3\degr$ P.A. and is shown
          using a filled ellipse in the lower right corner.  Contour
          levels are at $\pm$(3,4,5,6,7,8,9,10) times the rms noise. Locations
          of discrete sources detected by \citet{2000ApJ...544..686L} and in this image
          are marked with cross symbols. This image, and the two
          images that follow have been deconvolved. Errors
          due to sidelobes are not significant, as calculated and discussed in the text. In this image, and all subsequent images, light regions and solid contours indicate $-$ve features; dark regions and dashed contours $+$ve features. 
        }
    \label{nsrcmap1}
  \end{center}
\end{figure}
\begin{figure}
  \begin{center}
    \includegraphics[height=70mm, angle=0]{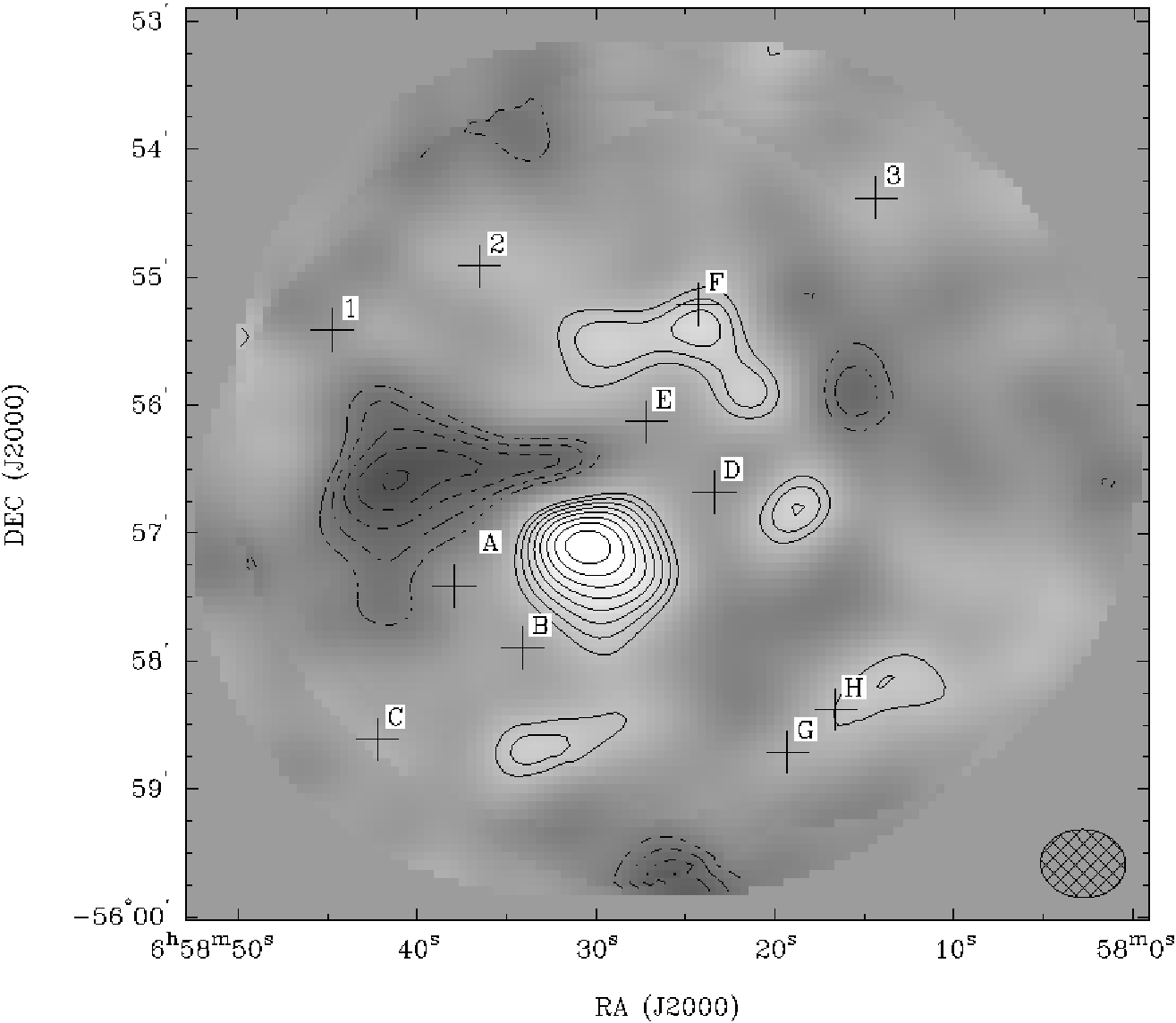}\\
    \caption
        {
          18 GHz mosaic image of the bullet cluster with unresolved continuum
          sources subtracted. The beam FWHM and rms noise are same as
          for figure \ref{nsrcmap1}. Contour levels are at $\pm$(3,4,5,6,7,8,9,10) times the rms noise. Locations of discrete sources
          detected by \citet{2000ApJ...544..686L} and in our 18 GHz data
          are marked with cross symbols.
      }
    \label{nsrcmap2}
  \end{center}
\end{figure}

\subsection{Imaging}
Multi-channel continuum visibilities in both bands and both pointings were imaged together to form a wide-field multi-frequency
synthesis mosaic image of the Bullet cluster shown in Fig.~\ref{nsrcmap1}.
The image displayed has been truncated where the attenuation in the mosaic
image owing to the primary beam---and consequently the response to source flux
density---falls below 10\%.   
The mosaic image was made with a beam of FWHM $20\farcs8 \times 14\farcs5$ at
$89\degr$ position angle (P.A.), and has an rms noise that is
5.5~$\mu$Jy~beam$^{-1}$ at the centre and increases by factor five to the edges of the field
of view.  The rms noise in the image was estimated from Stokes V data and is
consistent with expectations based on the system temperature during the
observations, bandwidth and integration time. The image displayed has been
tapered to make the rms noise uniform over the field of view.

Discrete sources detected by \citet{2000ApJ...544..686L} at 1.344~GHz are marked using
symbols `A' through `H'.
All these eight sources are detected in the 18~GHz mosaic image at a level
exceeding two standard deviation in image thermal noise.  
The brightest source in our 18-GHz image is at RA: $06^{\rm h}58^{\rm m}37\fs9$, 
DEC: $-55\degr57\arcmin25\arcsec$ and is marked `A' in the figure; this
is a LIRG high-z galaxy \citep{2009ApJ...703..348R} and is detected with
a flux density of $1.69\pm0.04$~mJy at 18~GHz.
\section{Imaging the Bullet cluster at 18 GHz}
\label{pts}
Radio images of the cluster field include discrete radio
sources, a radio halo \citep{2000ApJ...544..686L}, and SZE \citep{2009ApJ...701...42H,2010ApJ...716.1118P} that is expected to
appear as a decrement at the ATCA observing frequencies. 
For the characterization of the SZE decrement and the Radio Halo, it is essential to accurately
subtract all discrete sources. 
\subsection{Subtraction of unresolved continuum sources}
The image in Fig.~\ref{nsrcmap1} was made with visibility weighting close to natural in
order to have low rms noise and good surface brightness sensitivity. To characterize first the unresolved continuum sources in the field, we
constructed a separate image with uniform weighting of the visibilities that
has a factor of two greater rms noise but, importantly, de-emphasizes extended
halo emission and SZE. Sky regions that have image intensity exceeding four
times the rms noise in Fig.~\ref{nsrcmap1} were examined for unresolved
sources.  Apart from sources `A' through `H', three
sources were detected with flux density exceeding four times the rms noise in
the uniformly weighted image; these are marked `1' through `3' in Fig.~\ref{nsrcmap1}.
Of these, sources `1' and `2' are undetected in the 1.344~GHz image and are likely 
to have relatively flatter spectra.  The flux densities of the 11 unresolved
continuum sources in the field were estimated from the image made with uniform
weighting and subtracted from the visibility data.

A mosaic image of the cluster made with natural weighting
and using visibilities from which unresolved continuum sources were subtracted
is in Fig.~\ref{nsrcmap2}. The beam FWHM and rms noise distribution and taper
are same as for Fig.~\ref{nsrcmap1}. 
\begin{center}
\begin{table*}
\caption{A comparison of the SZ feature properties before and after deconvolution.}
\label{beforeafterdeconv}
\begin{tabular}{ccccccccccc}
\hline\hline
Label & RMS$_{\rm NOISE}$ & Intensity & Significance & Size & \multicolumn{6}{c}{RA (J2000) DEC} \\ \hline 
 & ($\mu$Jy beam$^{-1}$) & ($\mu$Jy beam$^{-1}$) & ($\sigma$) & (3$\sigma$ contour) & h & m & s & $^\circ$ & $\arcmin$ & $\arcsec$ \\ 
\hline
\hline
Before & 7.0 &$-$93.4 & 10.5 & 56$\arcsec\times$81$\arcsec$ & 06 & 58 & 30.762 & $-$55 & 57 & 08  \\
After  & 6.5 & $-$84.7 & 10.7 & 66$\arcsec\times$76$\arcsec$ & 06 & 58 & 30.285 & $-$55 & 57 & 08   \\ 
\hline\hline
\end{tabular}
\end{table*}
\end{center}
\begin{figure}
  \begin{center}
    \includegraphics[height=70mm, angle=0]{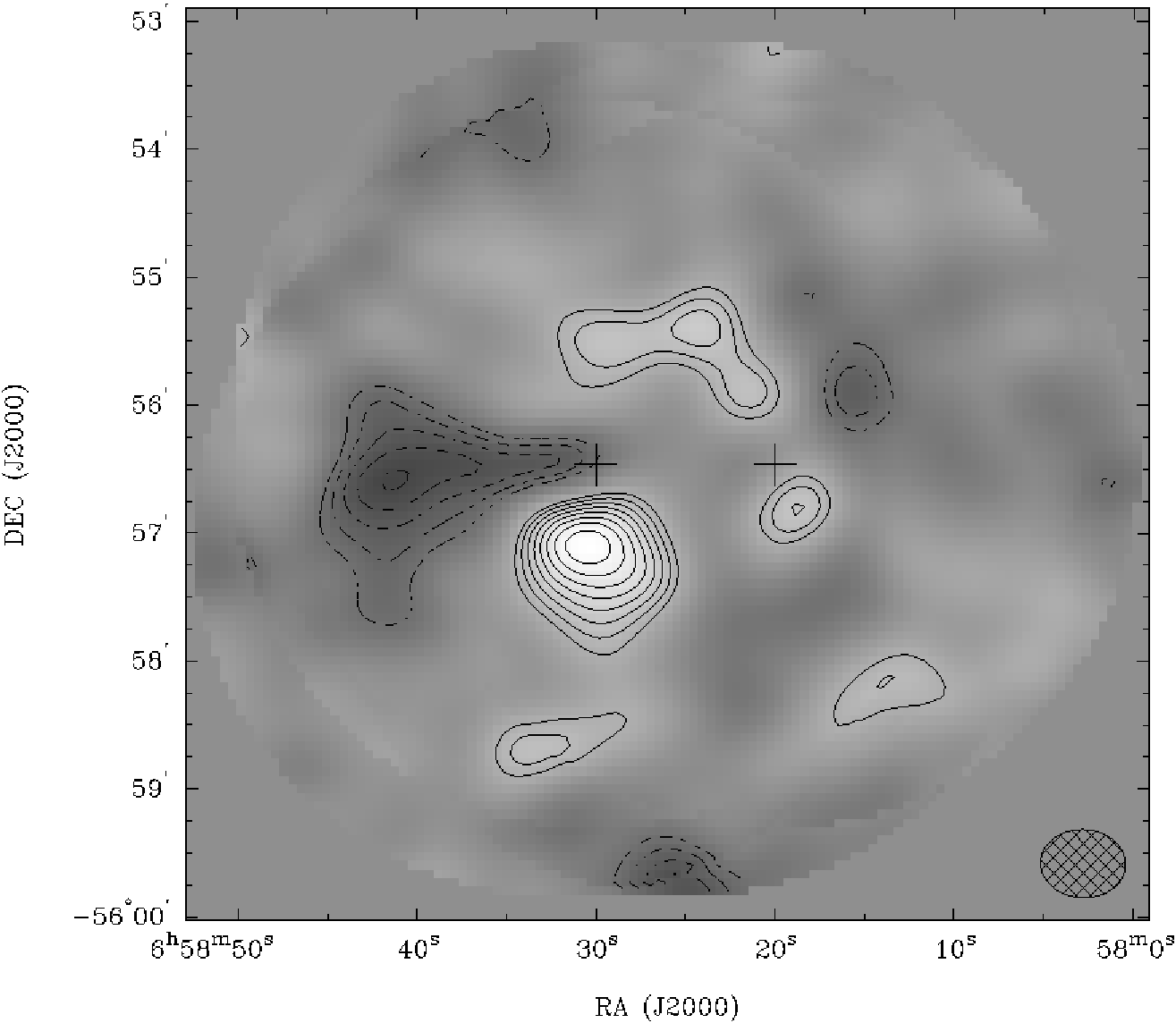}\\
    \caption
        {
          Source--subtracted deconvolved 18 GHz mosaic image of the bullet
          cluster (identical to the previous image), made with natural weighting and a uv taper
          corresponding to a $39\farcs8 \times 32\farcs1$ FWHM beam. Image rms noise level is
          6.5~$\mu$Jy~beam$^{-1}$. Contour levels are at $\pm$(3,4,5,6,7,8,9,10) times the rms noise. The radio
          halo previously detected by \citet{2000ApJ...544..686L} is
          the brightest positive feature close to the centre of the
          image and has a peak of 7.8$\sigma$; the SZ effect is detected and has a
          peak decrement of $-$10.7$\sigma$ to the S of the halo. Two
          other SZ effect features are detected to the NW regions of
          the halo, with a peak decrement of $-$5.7$\sigma$. Contrast has been adjusted in this image in order to make all contours visible. The two peaks
          seen in X--rays are marked. Notice the $\sim$40$\arcsec$ displacement of the peak SZE decrement from the X--ray brightness peak
      }
    \label{nos_pix130}
  \end{center}
\end{figure}
\begin{figure}
  \begin{center}
    \includegraphics[height=70mm, angle=0]{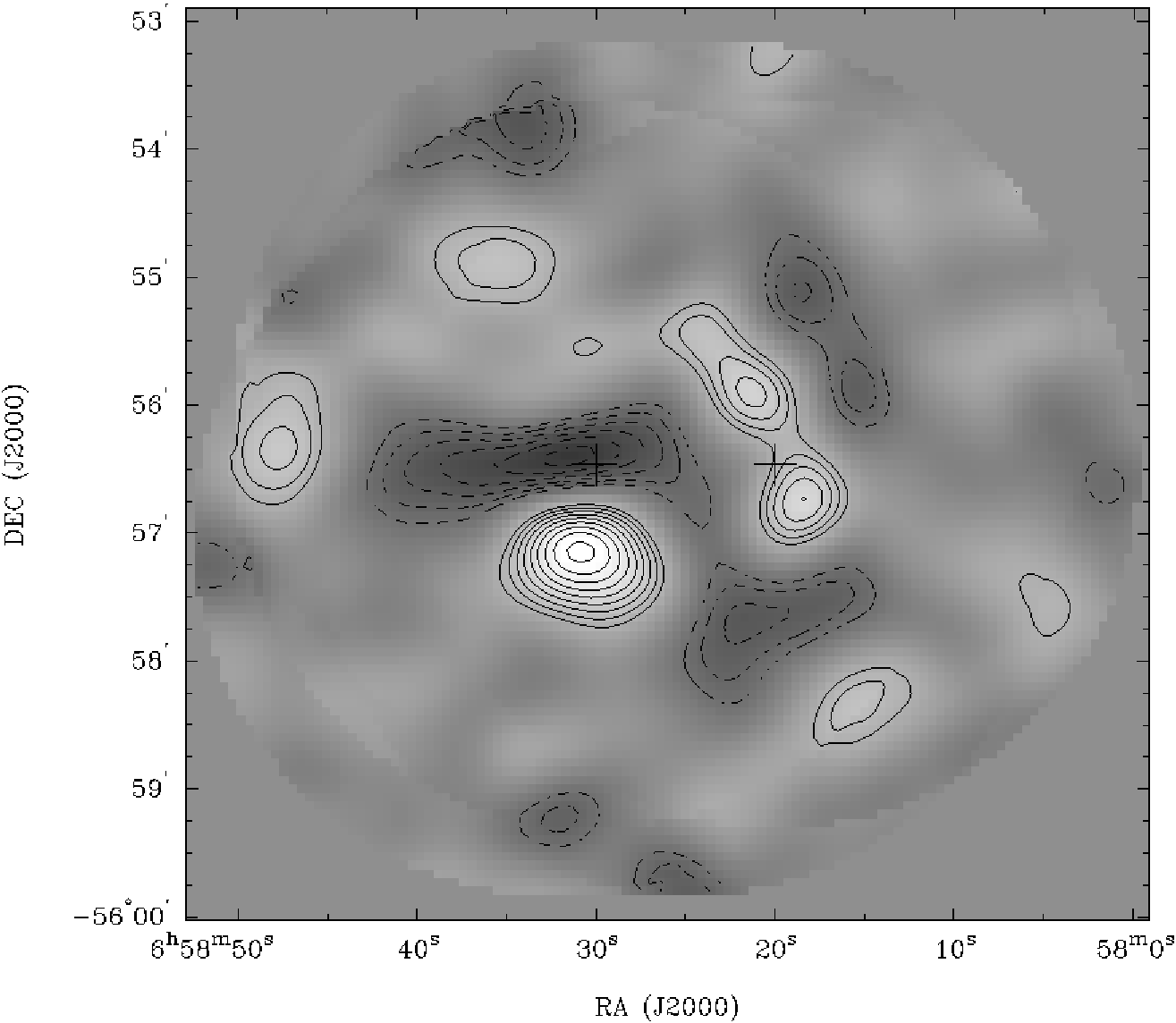}\\
    \caption
        {
          Source--subtracted non--deconvolved 18 GHz mosaic image of the bullet
          cluster, made with natural weighting and a uv taper
          corresponding to a $39\farcs8 \times 32\farcs1$ FWHM beam. Image rms noise level is
          6.5~$\mu$Jy~beam$^{-1}$. Contour levels are at $\pm$(3,4,5,6,7,8,9,10) times the rms noise. The radio
          halo previously detected by \citet{2000ApJ...544..686L} is
          the brightest positive feature close to the centre of the
          image and has a peak of 7.8$\sigma$; the SZ effect is detected and has a
          peak decrement of $-$10.7$\sigma$ to the S of the halo. Two
          other SZ effect features are detected to the NW regions of
          the halo, with a peak decrement of $-$5.7$\sigma$. Contrast has been adjusted in this image in order to make all contours visible. The two peaks
          seen in X--rays are marked. Notice the $\sim$40$\arcsec$ displacement of the peak SZE decrement from the X--ray brightness peak
      }
    \label{nos_pix130nondeconv}
  \end{center}
\end{figure}
\begin{figure}
  \begin{center}
    \includegraphics[height=70mm, angle=0]{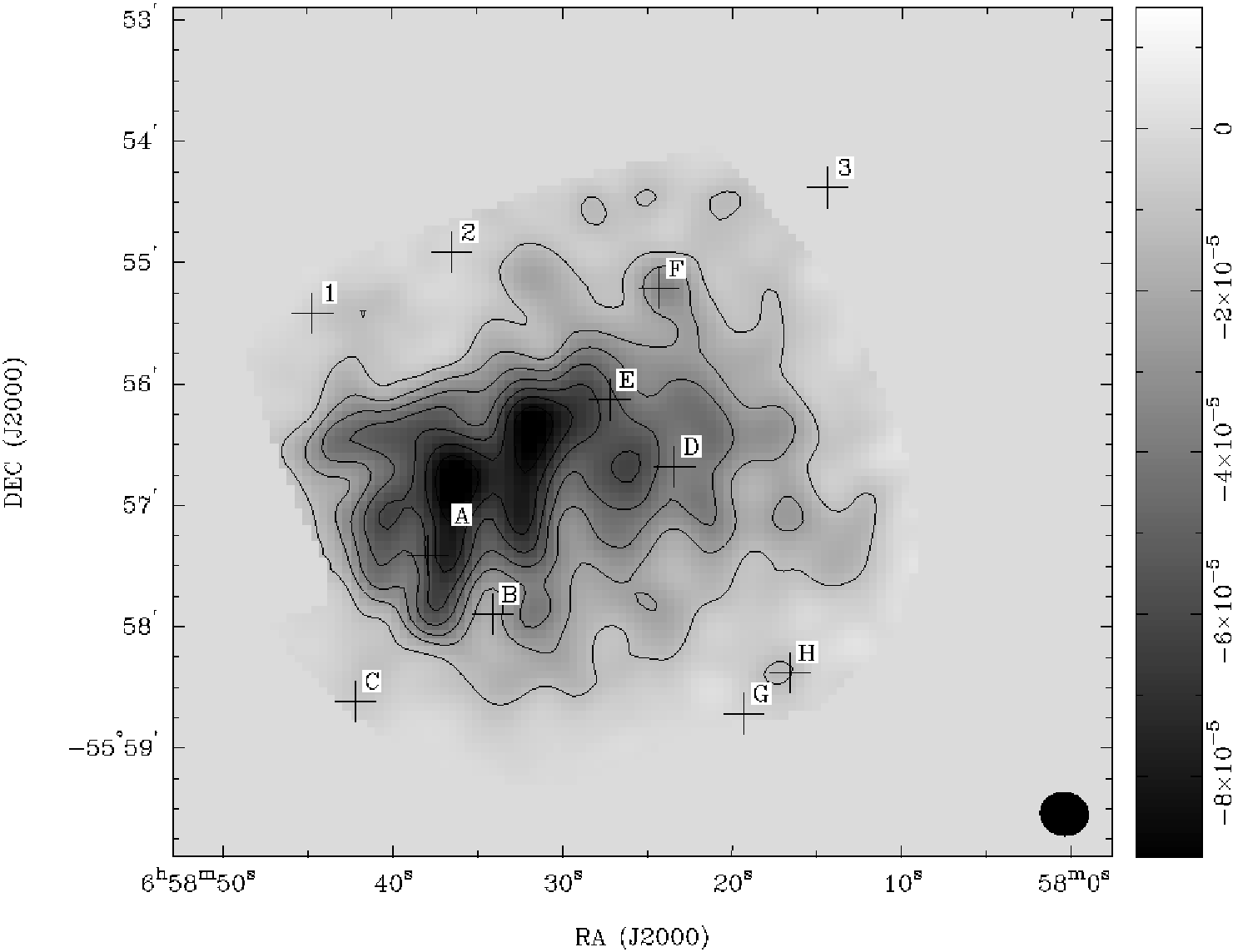}\\
    \caption
        {
          An image of the radio halo in the Bullet cluster. This image
          was obtained by convolving the 1.344 GHz radio halo image
          with the 18 GHz synthesized beam and then applying a gain taper for the
          primary beam at 18 GHz.
      }
    \label{halo1344spx}
  \end{center}
\end{figure}
Positions of all the sources subtracted have been marked in
Fig.~\ref{nsrcmap1} and in Table~\ref{ptsrc} we list the positions and 18-GHz flux
densities. A negative feature appears close to the location where
source `F' was subtracted; this is an extended negative feature that is absent in the
higher resolution images made with uniform visibility weighting. The brightest
positive feature in this residual image is extended emission with a peak of
40~$\mu$Jy~beam$^{-1}$ at RA: $06^{\rm h}58^{\rm m}31\fs1$, DEC: $-55\degr56\arcmin27\arcsec$;
this is 18-GHz emission from the radio halo, which was previously
discovered by \citet{2000ApJ...544..686L}. At RA: $06^{\rm h}58^{\rm
  m}31\fs7$, DEC: $-55\degr57\arcmin04\arcsec$ is the peak negative
feature; this is the centre of the SZ feature, discussed below.
\begin{center}
\begin{table}
\caption{Unresolved continuum radio sources detected in the Bullet cluster field.}
\label{ptsrc}
\begin{tabular}{cccccccr}
\hline\hline
Source & \multicolumn{6}{c}{RA (J2000) DEC} & {Flux
  Density} \\ \hline 
  & h & m & s & $^\circ$ & $\arcmin$ & $\arcsec$ & $S_{\rm 18~GHz}$ (mJy) \\ 
\hline
A & 06 & 58 & 38.0 & $-$55 & 57 & 25 & $1.69\pm0.04$ \\
B & 06 & 58 & 33.8 & $-$55 & 57 & 58 & $0.06\pm0.04$   \\
C & 06 & 58 & 41.4 & $-$55 & 58 & 41 & $1.4\pm0.4$ \\
D & 06 & 58 & 23.8 & $-$55 & 56 & 41 & $0.07\pm0.02$  \\
E & 06 & 58 & 27.5 & $-$55 & 56 & 06 & $0.05\pm0.02$ \\
F & 06 & 58 & 24.5 & $-$55 & 55 & 13 & $0.11\pm0.03$   \\
G & 06 & 58 & 19.3 & $-$55 & 58 & 41 & $0.26\pm0.12$  \\
H & 06 & 58 & 16.1 & $-$55 & 58 & 24 & $0.13\pm0.08$ \\
1 & 06 & 58 & 44.6 & $-$55 & 55 & 24 & $0.5\pm0.2$ \\
2 & 06 & 58 & 36.7 & $-$55 & 54 & 57 & $0.18\pm0.06$  \\
3 & 06 & 58 & 14.6 & $-$55 & 54 & 23 & $0.53\pm0.13$  \\ \hline\hline
\end{tabular}
\begin{tablenotes}[flushleft]
\item {\sc notes}-- Letters indicate discrete sources listed in \citet{2000ApJ...544..686L}
  and have the same letters as in their Table~2; numbers indicate additional discrete
  sources from our 18-GHz observations. 
\end{tablenotes}
\end{table}
\end{center}
\subsection{The 18 GHz Bullet cluster image: Radio Halo and the SZ Effect}
\label{halosub1}
The 18-GHz image in Fig.~\ref{nos_pix130}, which has been made with a taper that represents a $39\farcs8 \times 32\farcs1$ FWHM beam, clearly shows negative intensity values towards the central regions of the cluster.  These SZE features appear to be confined
within, and are bounded by the lowest intensity contours of the Radio Halo component, as can be seen in the 18-GHz image (Fig.~\ref{nos_pix130}) prior to any subtraction of the halo component, which would only enhance these SZE features. 
In this image, which has the unresolved continuum sources subtracted,
there is a peak in flux density at the peak of the Radio Halo and positive 18-GHz emission
towards the eastern parts of the Radio Halo. Most noteworthy is the negative peak,
which may be a deep SZE feature, to the south of the peak in the
expected halo contribution, at RA: $06^{\rm h}58^{\rm m}30\fs285$, DEC: $-55\degr57\arcmin08\arcsec$. It may be noted that this interferometer image lacks the low spatial frequencies to reproduce the extended SZE emission, which may be missing.

The 18 GHz image made from ATCA (Fig.~\ref{nos_pix130}) exhibits not just the SZE (peak
decrement $-84.7\mu$Jy beam$^{-1}$ at $-$10.7$\sigma$), but also, surprisingly, the radio halo (peak 7.8$\sigma$). Radio halos are known to have spectra that steepen at frequencies less than 5 GHz. The spectral index of the radio halo in the Bullet cluster does not show any sign of steepening at frequencies lower than 8 GHz \citep{2000ApJ...544..686L}, and the apparent steepening at 8.8 GHz may well be due to the SZE. To the best of our knowledge, this is the first positive detection of a radio halo at any frequency above 10 GHz; further discussion of the radio halo at 18 GHz and its
implications is left to a companion manuscript.
Therefore, in conclusion, the SZE shown in Fig.~\ref{nos_pix130} is the {\em least} 
that can be expected at 18 GHz, with a peak SZE dip of
$-$10.7$\sigma$.

It is possible to compare the derived Compton--y parameter from
our observations and compare it with the derived Compton--y from the
APEX--SZ observations of \citet{2009ApJ...701...42H} even though
they do not derive a Compton--y from the peak SZE they observe. 
We estimate it from the peaks of the SZE in the two observations -- those presented in this paper and
those in \citet{2009ApJ...701...42H} -- and compare them. The Compton--y parameter from our data (without radio halo subtraction) is
$y$=7.0$\times$10$^{-5}$ whereas the Compton--y from
\citet{2009ApJ...701...42H}'s observations is
$y$=2.95$\times$10$^{-4}$. Our 18~GHz ATCA image of the SZE distribution is an interferometer image and would miss larger scale extended
features.  The smallest projected baselines in the ATCA observations correspond
to the antenna diameter of 22~m, and the largest angular scale structures reproducible in the ATCA image would be less than the antenna beam FWHM
of $2\farcm6$.  Because of missing large-scale SZE structure the absolute depth of
SZE features might be reduced.

The discrepancy between the Compton--y computed in the ATCA image compared to the APEX image shows that the SZE is not just the compact features detected in this work, and halo subtraction and sensitivity to larger scale SZE features are important to image the SZE in this cluster in its entirety.

In the next section, we describe consistency checks on our
data and 18 GHz images, to prove that the SZ feature discussed above
is not an artifact of improper deconvolution or incorrect
point--source removal. 
\section{Tests of robustness / Error Estimates}
\label{consist}
\begin{center}
\begin{table*}
\caption{A comparison of the compact SZ feature properties before and after point--source subtraction.}
\label{beforeafterptsrc}
\begin{tabular}{ccccccccccc}
\hline\hline
Label & RMS$_{\rm NOISE}$ & Intensity & Significance & Size & \multicolumn{6}{c}{RA (J2000) DEC} \\ \hline 
 & ($\mu$Jy beam$^{-1}$) & ($\mu$Jy beam$^{-1}$) & ($\sigma$) & (3$\sigma$ contour) & h & m & s & $^\circ$ & $\arcmin$ & $\arcsec$ \\ 
\hline
Before & 6.5 & 35.2 & 5.8 & 64$\arcsec\times$56$\arcsec$ & 06 & 58 & 30.285 & $-$55 & 57 & 08  \\
After  & 6.5 & 84.7 & 10.7 & 64$\arcsec\times$56$\arcsec$ & 06 & 58 & 30.285 & $-$55 & 57 & 08   \\ 
\hline\hline
\end{tabular}
\begin{tablenotes}[flushleft]
\item {\sc notes}-- The shape of the SZ feature remains unchanged; positional differences between 3$\sigma$ contours of the SZ feature before and after point--source subtraction are $<$1$\arcsec$.
\end{tablenotes}
\end{table*}
\end{center}
\subsection{Error Estimates for point--source residuals}
\begin{center}
\begin{table*}
\caption{Sidelobe levels at the SZE and Radio Halo feature peaks}
\label{sidelobes}
\begin{tabular}{lrrrrrrr}
\hline\hline
Source &\multicolumn{2}{r}{Distance ($\arcsec$) from} &\multicolumn{2}{r}{Sidelobe levels at} & Residual &\multicolumn{2}{r}{Residual$\times$Sidelobe}\\ 
\hline 
            & RHM & SZM & RHM & SZM & ($\sigma$) & RHM ($\sigma$)& SZM ($\sigma$)\\ 
\hline
A & 56  & 53 & 0.0068 & $-$0.0096 & 0.5 & 0.0340 & $-$0.0480 \\
B & 103 & 56 & $-$0.0086 & $-$0.0370 & $-$0.6 & 0.0052& 0.0222\\
C & 125 & 124 & 0.0100 & $-$0.0140 & $-$1.3 & $-$0.0130&0.0182 \\
D & 146 & 79 & 0.0015 & $-$0.0030 & 0.2 & 0.0003&$-$0.0006\\
E & 119 & 70 & $-$0.01600 & $-$0.0145 & $-$0.3 &0.0048 & 0.0043\\
F & 145 & 120 & $-$0.0050 & 0.0035 & $-$2.0 & 0.0200&$-$0.0140\\
G & 160 & 147 & 0.0090 & 0.0135 & $-$1.6 &$-$0.0144 &$-$0.0216 \\
H & 236 & 158 & 0.0068 & 0.0130 & $-$3.0 &$-$0.0204 &$-$0.0390 \\
1 & 77  & 143 & $-$0.0063 & 0.0170 & 0.5 &$-$0.0032 &0.0085 \\
2 & 106 & 131 & 0.0130 & 0.0020 & $-$2.0&$-$0.0260 &$-$0.0400 \\
3 & 252 & 212 & 0.0080 & $-$0.0065 & $-$2.0&$-$0.0160 &0.0130 \\
\hline
Sum: & & &$+$0.0192 &$-$0.0426 &  &$-$0.0593&$-$0.0538 \\
\hline
\hline
\end{tabular}
\begin{tablenotes}[flushleft]
\item {\sc notes}-- In this table, as later, we denote the peak SZE dip position by `SZM' and the peak of the radio halo by
`RHM'. 
\end{tablenotes}
\end{table*}
\end{center}
\subsubsection{Point--source subtraction error estimate} 
Sidelobe level at the SZE peak dip and the Radio Halo peak sites due to each unresolved continuum source was
estimated, from the beam pattern corresponding to the image in
Fig. \ref{haloxray2}. These values are stated as fractions in Table \ref{sidelobes}, in columns 4 and 5.
Next, the residuals at the point--source sites are estimated; these are listed in column 6 in Table \ref{sidelobes} as fractions of the noise rms. Estimates of the errors caused due to sidelobes of these residuals were then estimated; these are listed in the last two columns in the same table. A worst case error in flux density at the position of the hole
is estimated by summing these error contributions; the value of this error estimate is less than 0.06$\sigma$ or
0.38$\mu$Jy beam$^{-1}$, for both the SZE peak dip and the RH peak. The contribution to uncertainty from point-source
subtraction is thus not significant. 
\subsection{Comparison of SZ feature before and after point--source
  subtraction}
\label{beforeafter}
Fig.(\ref{nsrcmap1}) shows the 18 GHz image with all point--sources
present; this image has a --ve feature peaking at RA: $06^{\rm h}58^{\rm m}30\fs285$, DEC: $-55\degr57\arcmin08\arcsec$
at $-$11.1$\sigma$ ($-$87.5$\mu$Jy beam$^{-1}$; noise RMS = 6.5$\mu$Jy
beam$^{-1}$). After modeling and removing all point--sources, this
--ve feature is still present, with its peak still at at RA: $06^{\rm
  h}58^{\rm m}30\fs285$, DEC: $-55\degr57\arcmin08\arcsec$, and a
peak of $-$10.7$\sigma$ ($-$84.7$\mu$Jy beam$^{-1}$; noise RMS = 6.5$\mu$Jy beam$^{-1}$, i.e. a difference of 2.8$\mu$Jy beam$^{-1}$). More importantly, the size, extent and shape of this --ve feature remain exactly the same after
point--source subtraction. We summarize the properties of the
--ve (SZ) feature before and after point--source subtraction in Table
\ref{beforeafterptsrc}. From the table, we may conclude the
following errors/limits on the properties of the the SZE
feature:
\begin{enumerate}
\item Error in Intensity: $\pm$2.8$\mu$Jy beam$^{-1}$
\item Error in Significance: $\pm$0.4$\sigma$
\item Error in Position: $\pm$0$\arcsec$
\item Error in Size: $<$1$\arcsec$
\end{enumerate}
\begin{figure}
  \begin{center}
    \includegraphics[height=70mm, angle=0]{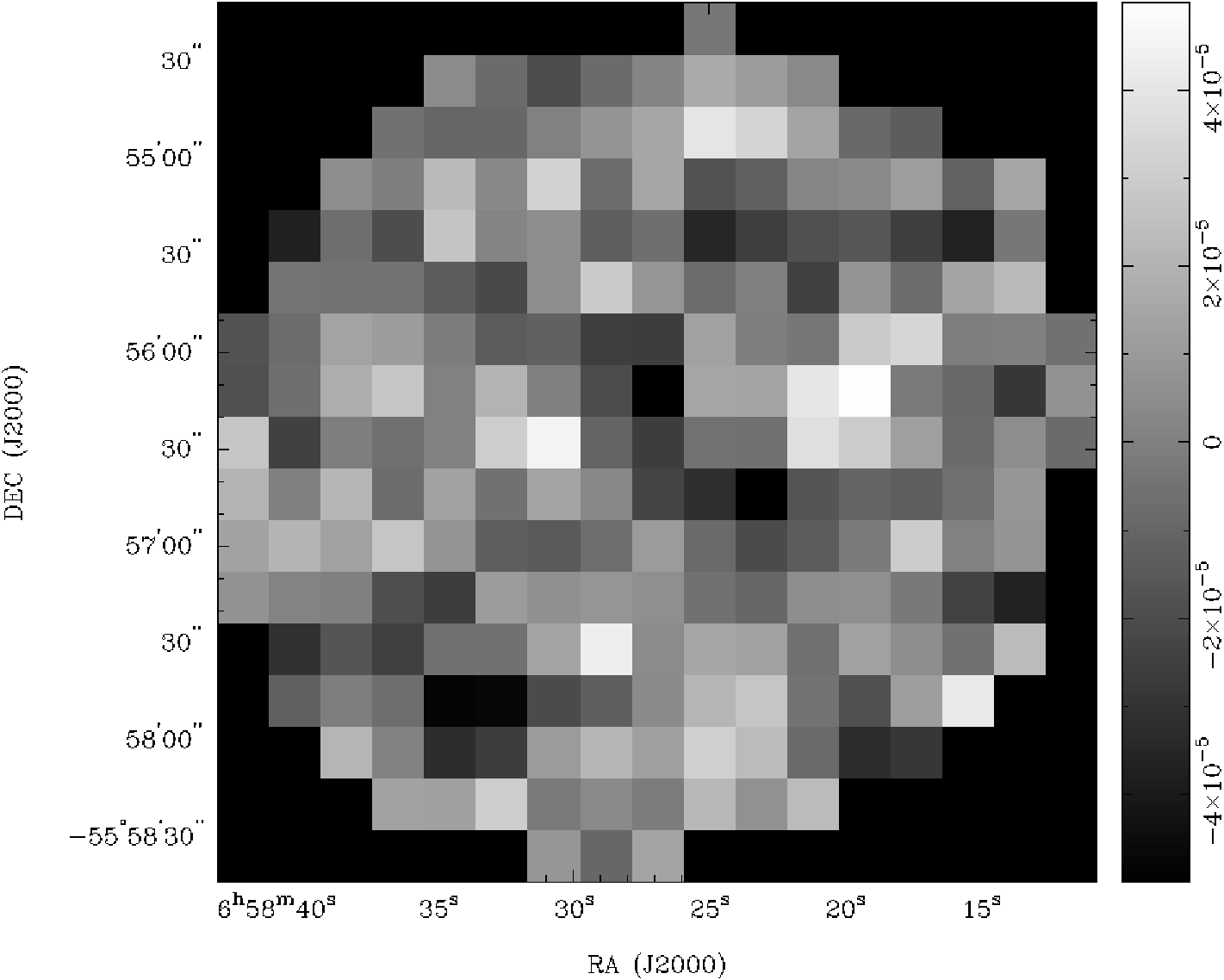}\\
    \caption
        {
          The difference between images made with the 2009 and 2010
          datasets. Noise RMS is 20$\mu$Jy beam$^{-1}$, and this
          difference image is consistent with gaussian noise.
      }
    \label{2009minus2010_16by16}
  \end{center}
\end{figure}
\subsection{Comparing 2009 and 2010 datasets}
\label{yearcomparison}
Our 18 GHz data was collected in two different arrays of the ATCA:
H168 and H75, in 2009 and 2010. Both the 2009 and 2010 datasets
contain data from both the arrays. We can therefore test the validity
of the features in the 18 GHz image by comparing images from the 2009
and 2010 datasets. A difference of the images made from the two
datasets was taken -- these images were made with a resolution of 16
$\arcsec$, to ensure that the pixels are independent data samples. The
difference is shown in Fig.(\ref{2009minus2010_16by16}). This
difference image was compared with a gaussian with RMS 20$\mu$Jy
beam$^{-1}$, using the non--parametric Mann--Whitney Wilcoxon test,
and it was found that z--value=0.7 and 2--tailed p--value=0.43. The
difference image is therefore consistent with gaussian noise.

This means that the features are not a result of phase errors in calibration.
\subsection{Comparing the shape and size of the SZ effect at 18 GHz to
  the 90 GHz APEX image}
\label{shapecompare}
The shape and size of the SZ effect in our 18 GHz may be compared to
the 90 GHz APEX single--dish image, provided that a reliable model of
the radio halo at 18 GHz exists. It is not possible to separate the
radio halo and the SZ given observations at a single frequency, but a
few reasonable assumptions may be made about the radio halo, in order
to estimate the SZ at 18 GHz. These are described in
\S\ref{haloretain} and \S\ref{2phase}. The resulting 18 GHz SZ images
were convolved to a 90 $\arcsec$ beam, as shown in
Figs.(\ref{haloxray3}\&\ref{haloxray5}). These are qualitatively
similar to Fig.(3) in \citep{2009ApJ...701...42H}. This is discussed
in further detail in \S\ref{haloretain}.
\section{Limits on the SZ Effect from 18 GHz ATCA image}
\label{ruleout}
In this section we enumerate and examine different possibilities for
the properties of the radio halo and the SZE and provide upper and lower limits on the peak of the SZE
from our observations, combined with those made by
\citet{2000ApJ...544..686L}. Our ATCA image misses the largest
scales (i.e. lowest uv--spacings), and all discussion below pertains to
the angular scales covered by our image.

We first made an image of the Radio Halo from
\citet{2000ApJ...544..686L}, extrapolated to 18 GHz, using the
spectral index quoted by them. In order to ensure the same
uv--coverage, we we convolved the \citet{2000ApJ...544..686L} image
with the synthesized beam at 18 GHz. We then weighed this image with a
taper representing the primary beam of the ATCA at 18 GHz. This
results in a radio halo image that has the same uv--coverage as our 18
GHz image, and is weighed with a taper representing the beam at 18
GHz. This image can now be used to put limits on the SZE at 18 GHz, as
described below.

In what follows, we denote the peak SZE dip position by `SZM' and the peak of the radio halo by
`RHM'. We enumerate and discuss the possibilities for radio halo
properties below, assuming that the radio halo has a
spectral index of -1.3 between 1.344 and 6 GHz, as measured by
\citet{2000ApJ...544..686L}. Any spectral index variations mentioned in the
following discussion are between 6 and 18 GHz.
\begin{figure}
  \begin{center}
    \includegraphics[height=70mm, angle=0]{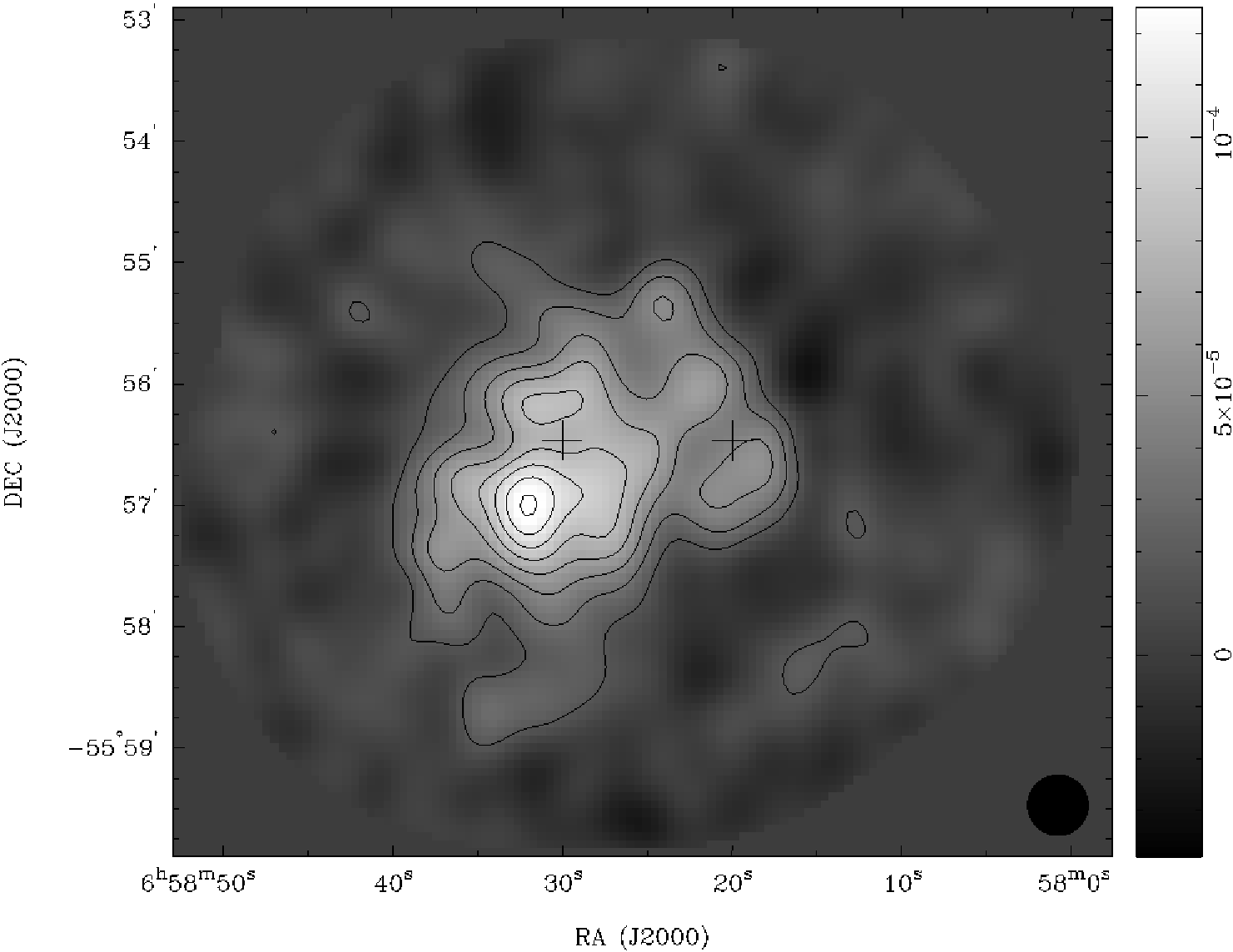}\\
    \caption{
       18 GHz mosaic image of the SZE in
      the Bullet cluster, with the radio halo subtracted assuming
      $\alpha=-$1.3 between 1.344 and 18 GHz, as measured by \citep{2000ApJ...544..686L}; the image has been smoothed to a beam of FWHM
      $30\arcsec$. Contours are at ($-$2,$-$4,$-$6,$-$8,$-$10,$-$12,$-$14,$-$16)
      times the image rms noise of 8.0~$\mu$Jy~beam$^{-1}$. Brightness
      centres of X--ray emission are marked.
    }
    \label{nohc30}
  \end{center}
\end{figure}
\begin{figure}
  \begin{center}
    \includegraphics[height=70mm, angle=0]{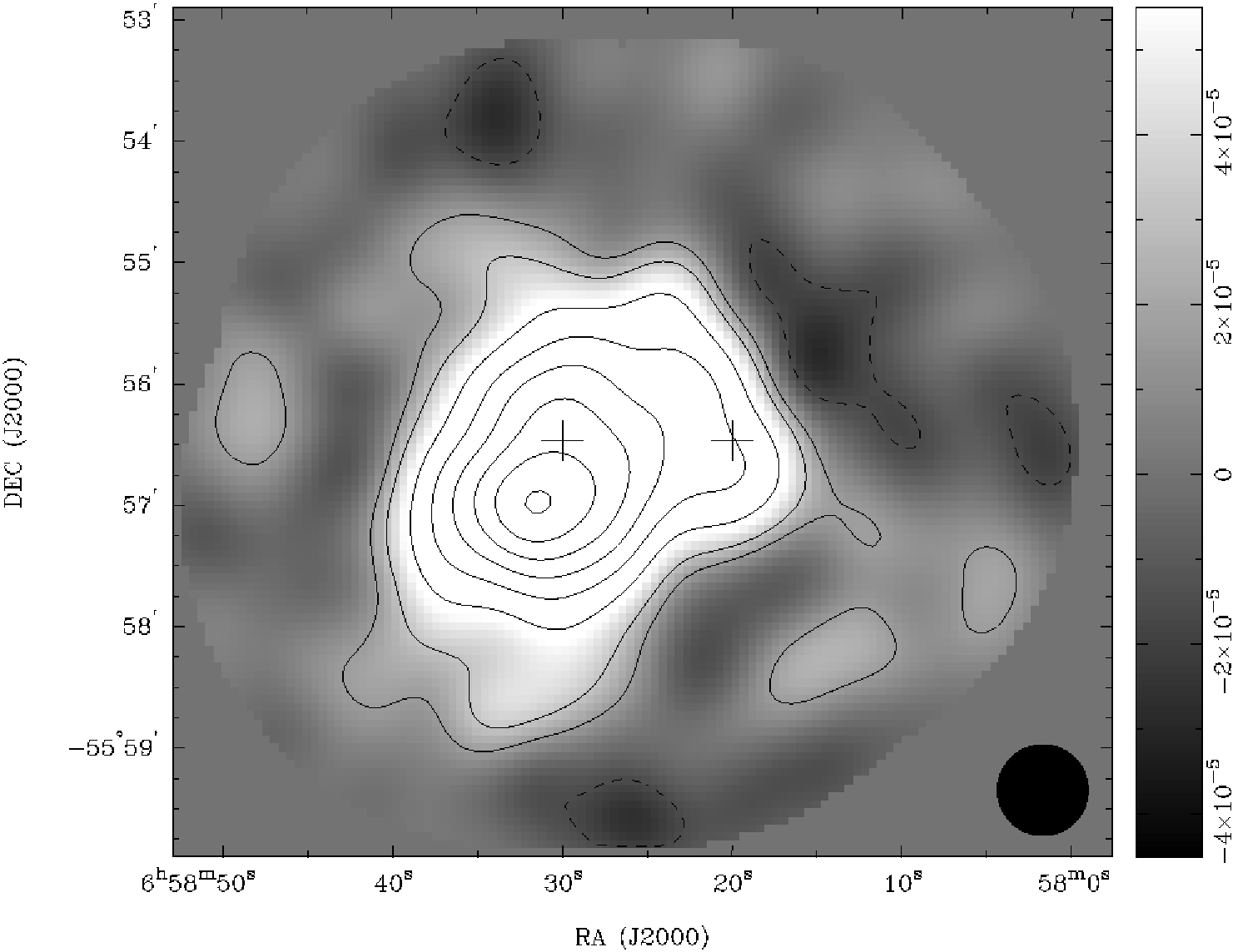}\\
    \caption{
      18 GHz mosaic image of the SZE in
      the Bullet cluster, with the radio halo subtracted assuming
      $\alpha=-$1.3, as measured by \citep{2000ApJ...544..686L}; the image has been smoothed to a beam of FWHM
      $45\arcsec$. Contours are at
      ($-$28,$-$24,$-$20,$-$16,$-$14,$-$12,$-$8,$-$4,$-$2,2)
      times the image rms noise of 8.0~$\mu$Jy~beam$^{-1}$. Brightness
      centres of X--ray emission are marked, as in the previous image.
    }
    \label{haloxray2}
  \end{center}
\end{figure}
\begin{figure}
  \begin{center}
    \includegraphics[height=70mm, angle=0]{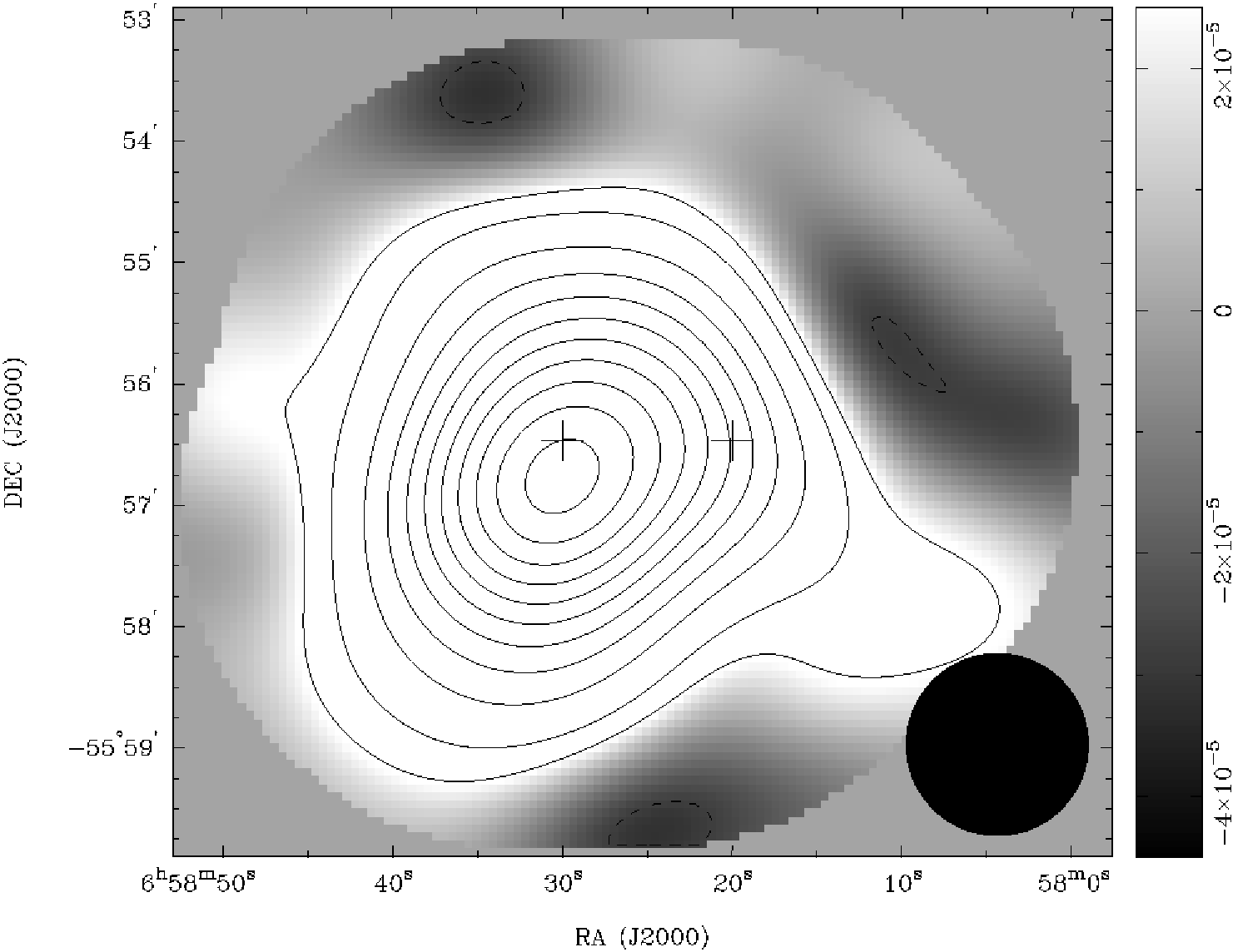}\\
    \caption{
       18 GHz mosaic image of the SZE in
      the Bullet cluster, with the radio halo subtracted assuming
      $\alpha=-$1.3, as measured by \citep{2000ApJ...544..686L}; the image has been smoothed to a beam of FWHM
      $90\arcsec$. Contours are at (2,$-$2,$-$4,$-$8,$-$12,$-$16,$-$20,$-$24,$-$28,$-$32,$-$36,$-$40)
      times the image rms noise of 14.0~$\mu$Jy~beam$^{-1}$. Brightness
      centres of X--ray emission are marked, as in the previous image.
    }
    \label{haloxray3}
  \end{center}
\end{figure}
\begin{figure}
  \begin{center}
    \includegraphics[height=70mm, angle=0]{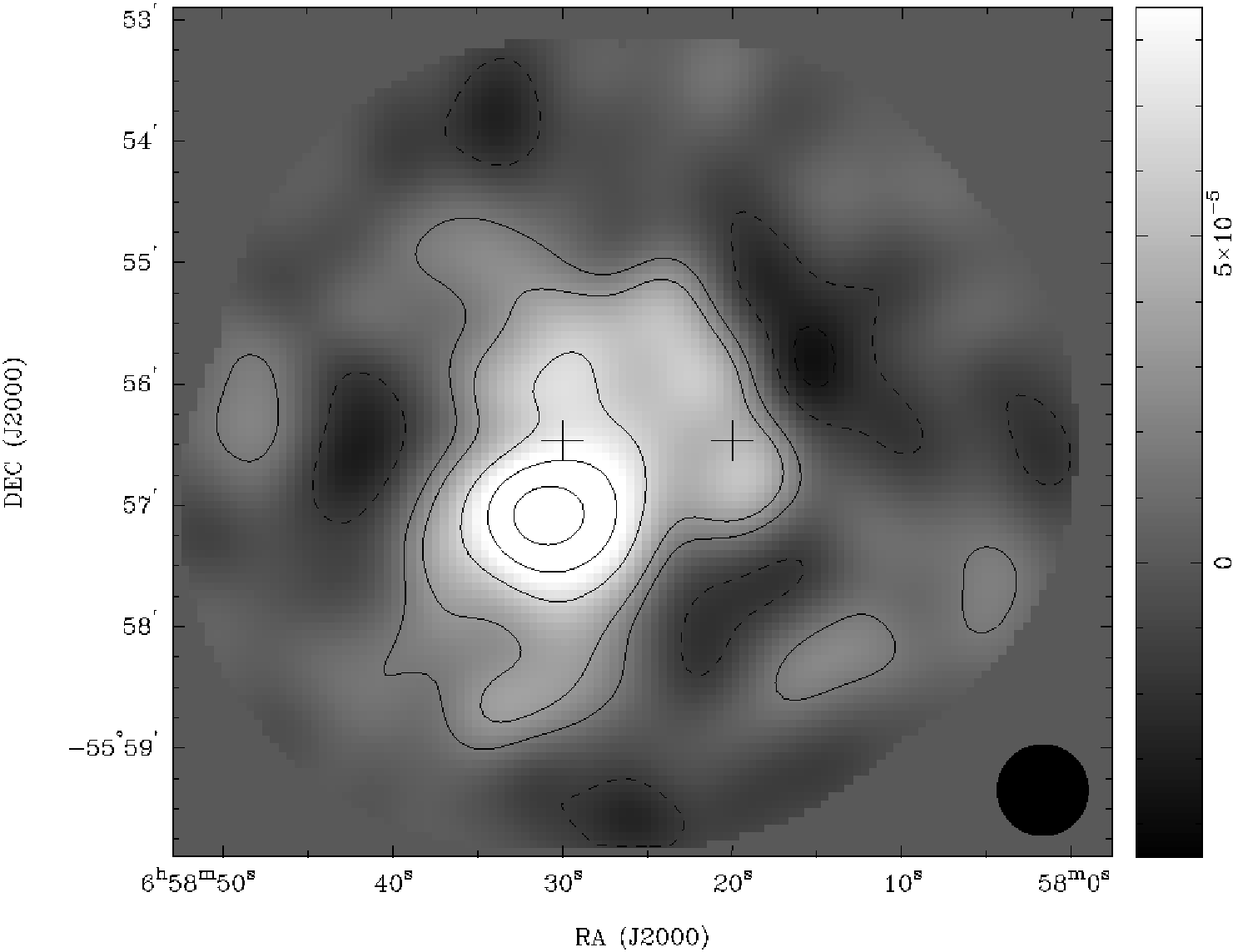}\\
    \caption{
       18 GHz mosaic image of the SZE in
      the Bullet cluster, with the radio halo subtracted assuming
      $\alpha=-$1.3 between 1.344 and 6 GHz and $\alpha=-$2.0 between
      6 and 18 GHz; the image has been smoothed to a beam of FWHM
      $45\arcsec$. Contours are at (2,$-$2,$-$4,$-$8,$-$12,$-$15)
      times the image rms noise of 8.0~$\mu$Jy~beam$^{-1}$. Brightness
      centres of X--ray emission are marked, as in the previous image.
    }
    \label{haloxray4}
  \end{center}
\end{figure}
\begin{figure}
  \begin{center}
    \includegraphics[height=70mm, angle=0]{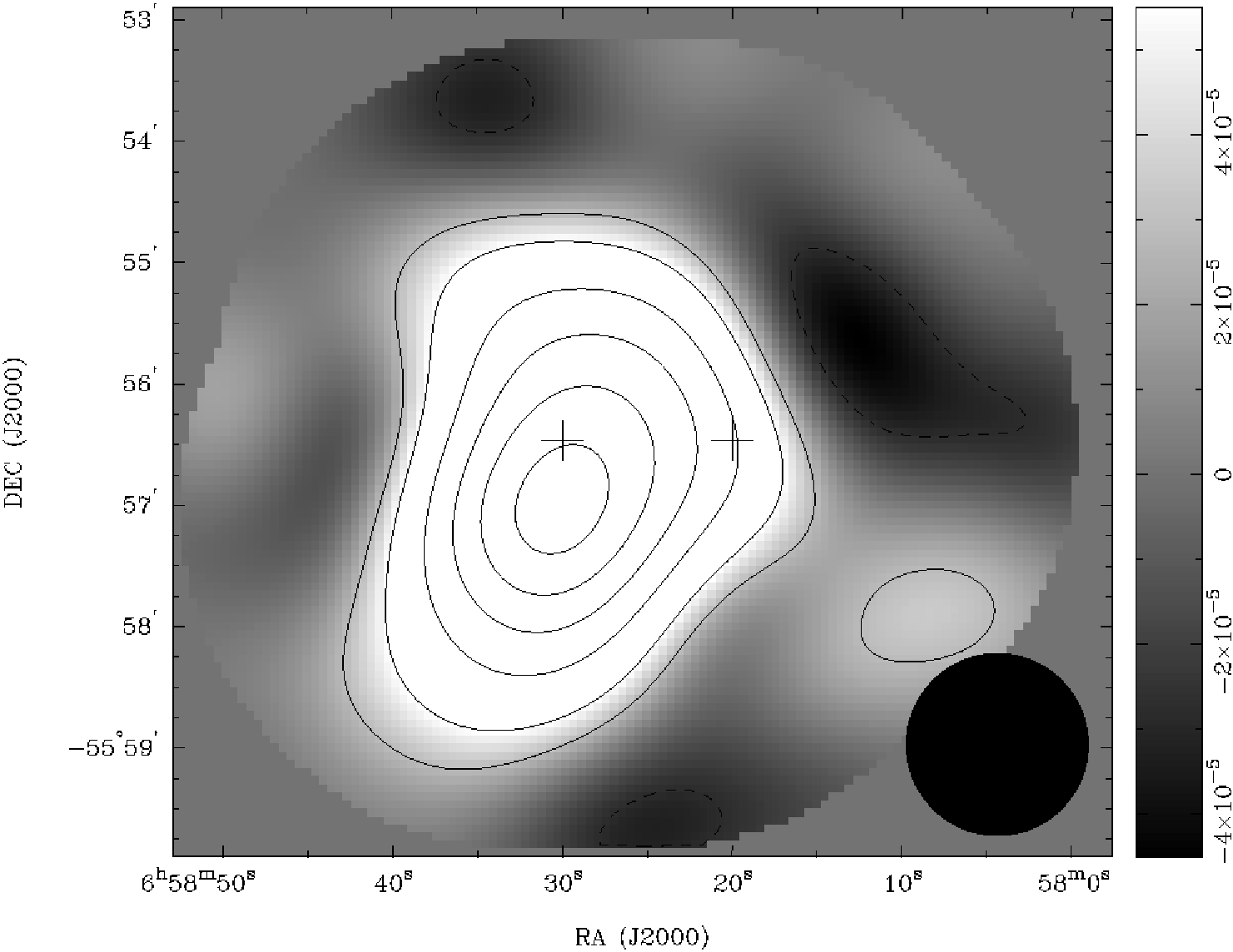}\\
    \caption{
       18 GHz mosaic image of the SZE in
      the Bullet cluster, with the radio halo subtracted assuming
      $\alpha=-$1.3 between 1.344 and 6 GHz and $\alpha=-$2.0 between
      6 and 18 GHz; the image has been smoothed to a beam of FWHM
      $90\arcsec$. Contours are at ($-$2,$-$4,$-$8,$-$11,$-$15,$-$20)
      times the image rms noise of 14.0~$\mu$Jy~beam$^{-1}$. Brightness
      centres of X--ray emission are marked, as in the previous image.
    }
    \label{haloxray5}
  \end{center}
\end{figure}
\subsection{Hypothesis I: The Radio Halo retains its properties up to 18 GHz}
\label{haloretain}
 The radio halo image we obtained, as described above, was multiplied
 the result with a factor that represents a spectral index of -1.3, the value quoted by \citet{2000ApJ...544..686L}.  We
subtracted this from our 18 GHz image, and resulting image is shown in
Fig.(\ref{nohc30}). We then convolved this image to 45$\arcsec$, obtaining the image
  shown in Fig.(\ref{haloxray2}). Brunetti et al. (2009) discuss
  spectral index variations at $\nu >$10 GHz, and from their
  discussions of primary, secondary and hybrid models of radio halo
  generation, it is unlikely that the radio halo in the Bullet cluster
  retains its spectral index and does not steepen at frequencies above 10 GHz. The image in
  Fig.(\ref{haloxray2}) is therefore an upper limit to the SZE in the
  Bullet cluster at 18 GHz, given the ATCA uv--coverage. 

  The peak SZE
  dip at SZM is $-$227$\mu$Jy beam$^{-1}$, which corresponds to a
  Compton--y of 1.7$\times$10$^{-4}$. Even after subtracting the
  maximum possible radio halo contribution and convolving with a
  45$\arcsec$ beam, the Compton--y at 18 GHz is less than that
  measured by the APEX single--dish instrument [REF]. This clearly
  indicates missing large--scale structure. However, the shape of the
  SZ effect at 18 GHz is similar to that of the SZ image at 90 GHz
  obtained from APEX. On convolving the radio halo subtracted image to
  90 $\arcsec$ (see Fig.(\ref{haloxray3})), the shape and size of the SZ image at 18 GHz
  qualitatively resembles the shape and size of the SZ image at 90
  GHz. This indicates that the missing large--scale structure in SZ is
  probably a smooth envelope that encompasses multiple compact features that are detected in the ATCA image and are spread over the SZE extent. In this scenario, our image detects substructure and this is distributed over the SZE extent suggesting that turbulence in gas pressure owing to the merger is widespread. However, it is well known that Fourier synthesis imaging of extended emission with high angular resolution tends to represent smooth structure as a patchy image arising from calibration errors and deconvolution instabilities. We present in Fig. \ref{nos_pix130nondeconv} a non--deconvolved source--subtracted image of the Bullet cluster, which is similar to the deconvolved image in Fig. \ref{nos_pix130}. 
  
 Additionally, we have compared two independent datasets in \S... above and confirmed that the substructure is identical within the expected errors associated with image noise. We have also shown in above that the contribution from point--source residuals to the error in the peaks of the radio halo and SZE  is insignificant; being a small fraction of the noise rms ($<$0.06$\sigma$).
\subsection{2--phase spectral index}
\label{2phase}
\citep{2011MNRAS.412..817B} have shown that in turbulent
(re--)acceleration models of radio halo formation, a steepening, or ``knee'' is expected
in the spectrum of the radio halo. The frequency at which this
``knee'' occurs is indicative of the age of the non--thermal plasma,
and the relative epoch of its (re--)acceleration, i.e. the epoch of the
cluster merger. Several radio halos known to be associated with
cluster mergers exhibit this steepening. The steepest known spectral
index for a radio halo is $\alpha$=2.0. As mentioned earlier, the
measurement of the spectral index of the radio halo in the Bullet
cluster is robust to atleast 6 GHz; therefore, we may build a simple
model of the radio halo spectral index, in which $\alpha$=1.3 from
1.344 to 6 GHz and $\alpha$=2.0 from 6 to 18 GHz. This model was then
subtracted from our 18 GHz image, and the subtraction followed the
same steps described in the previous sub--section. The result of this
subtraction, convolved with a 45$\arcsec$ beam, is shown in
Fig.(\ref{haloxray4}). The peak SZE
  dip at SZM is $-$145$\mu$Jy beam$^{-1}$, which corresponds to a
  Compton--y of 1.1$\times$10$^{-4}$. 
\begin{figure}
  \begin{center}
    \includegraphics[height=70mm, angle=0]{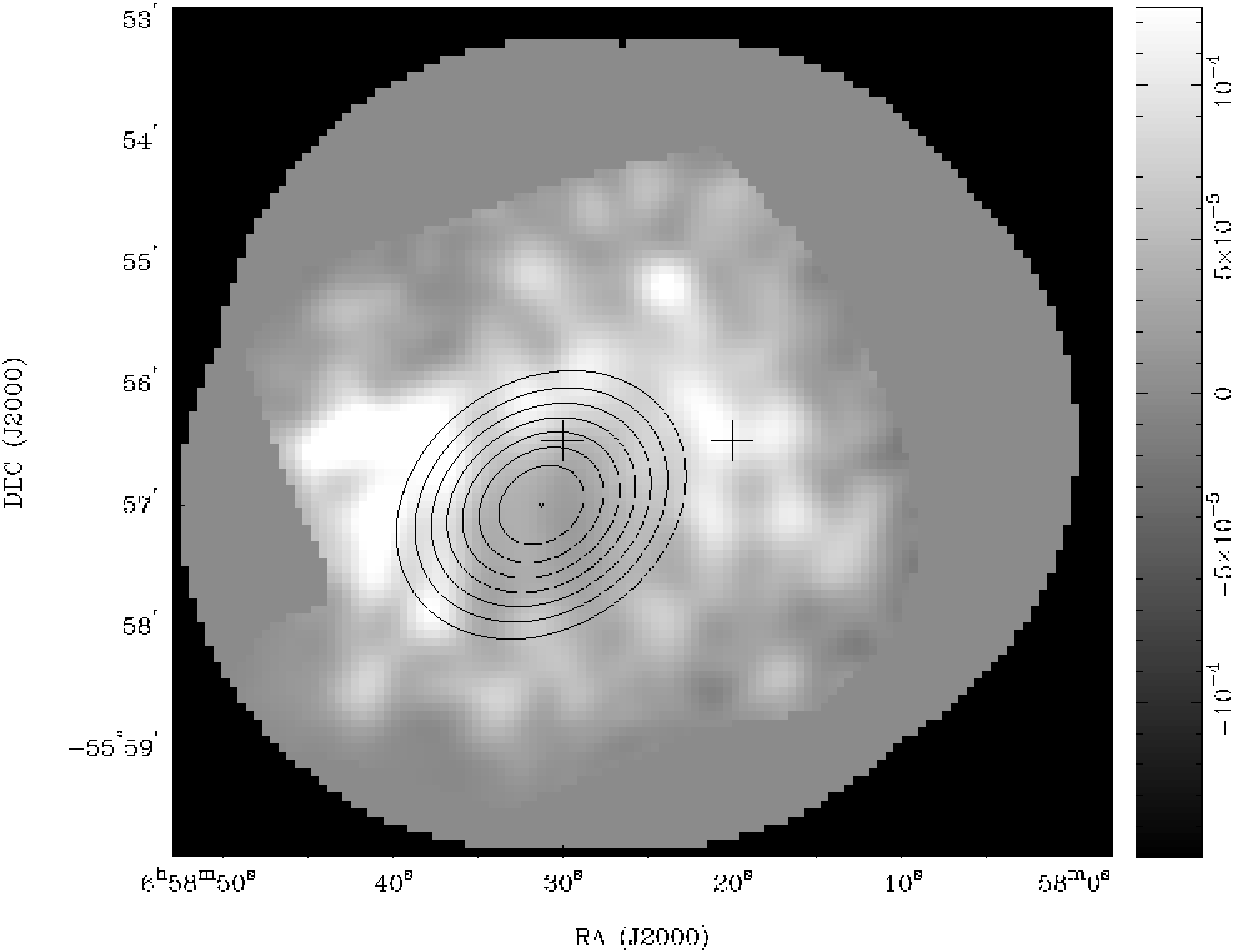}\\
    \caption
        {
         A model of the radio halo at 18 GHz, given the spectral index
         model which de--emphasizes the compact SZE feature, shown as
         a contour.
      }
    \label{halo_extreme1}
  \end{center}
\end{figure}
\begin{figure}
  \begin{center}
    \includegraphics[height=70mm, angle=0]{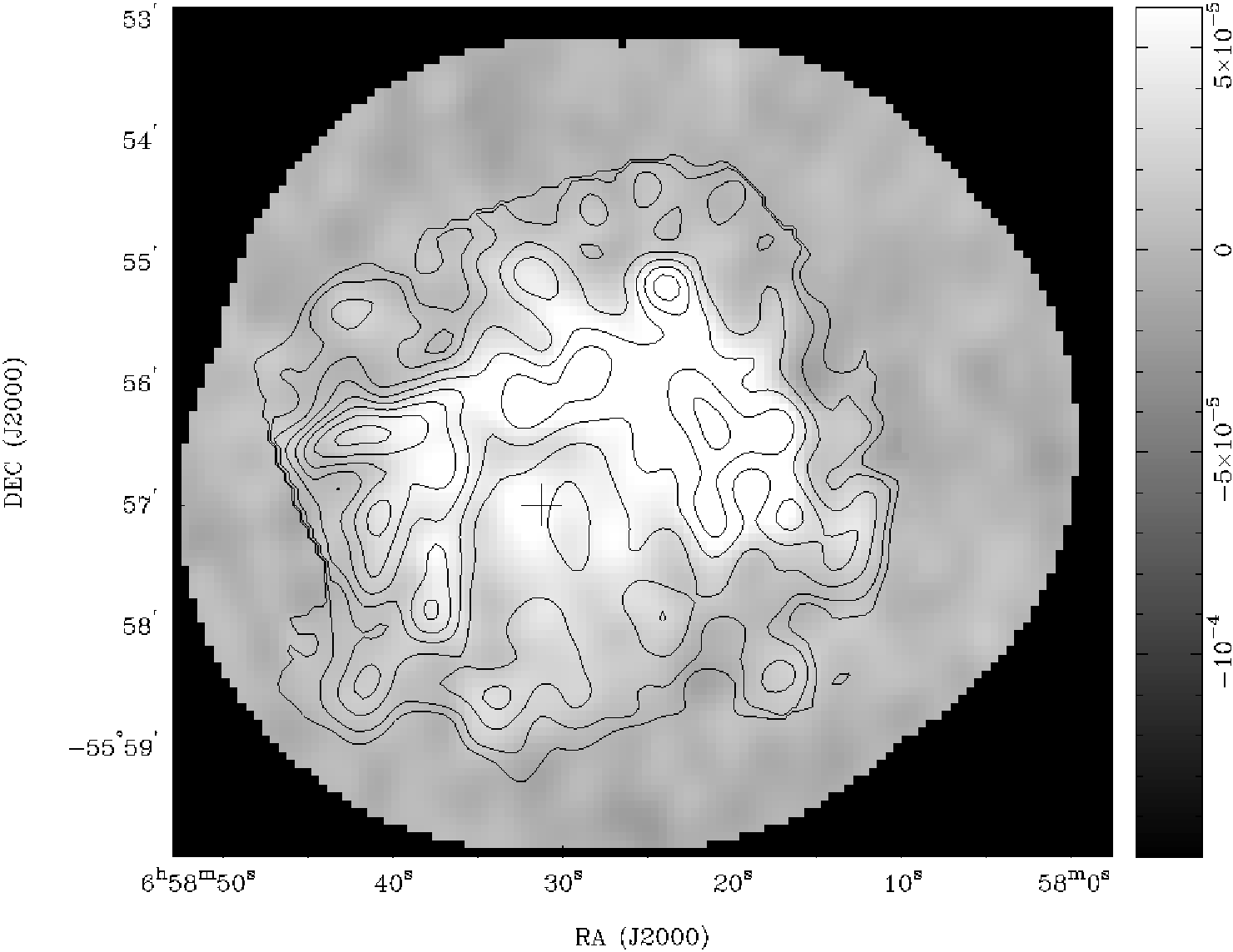}\\
    \caption
        {
          The result of subtracting the extreme halo in Fig.(\ref{halo_extreme1})
          above, with contours of the radio halo model at 18 GHz (shown in the
          previous figure). The subtraction clearly leads to several compact
          features, most of which are caused by the compact features
          in the radio halo model. However, the compact SZ feature
          seen in our original 18 GHz feature is still present. As
          discussed in the text, this implies that a simple spatial
          variation that de--emphasizes the radio halo at the SZM is not
          sufficient to make the compact SZE feature go away.
      }
    \label{halo_extreme_sub1}
  \end{center}
\end{figure}
\subsection{Spatial spectral index variations} 
  Though \citet{2000ApJ...544..686L} rule out spatial variations in the radio
  halo spectral index between 1.344 and 9 GHz, halo formation models
  do not rule out spatial variations in the spectral index above 10
  GHz. In general, a flat index at RHM and a steep spectrum at SZM would
  minimize the SZE. We therefore simulated such a spectral index which would
  de--emphasize the SZE, in order to explore the effect of extreme
  variations of the spectral index on the nature of the SZE, as
  follows.

  First, the compact feature in Fig.(\ref{nsrcmap2}) was modeled as a gaussian
  and its properties (RA, Dec, major axis, minor axis) noted. Then, a
  gaussian spectral index map (for the radio halo) was simulated centered at the position
  of the compact SZE feature in Fig.(\ref{nsrcmap2}), so that the index would be
  steepest at the centre of the gaussian derived above, and would
  flatten out in the outer regions of the gaussian. The value chosen at
  the centre was such that only $\sim$10\% of the radio halo would be
  present at the centre of the compact SZE feature at 18 GHz. This
  spectral index map is shown as a contour in Fig.(\ref{halo_extreme1}). The radio halo image,
  weighed by the appropriate taper, as described above, was then
  multiplied with a factor that represents a constant spectral index
  between 1.344 and 6 GHz, and a spectral index variation we simulated
  above, for frequencies between 6 and 18 GHz. The resulting image of
  the radio halo at 18 GHz is shown in Fig.(\ref{halo_extreme1}). This image
  represents the radio halo at 18 GHz, given the simulated spectral
  index, and was subtracted from our 18 GHz image, resulting in the
  image shown in Fig.(\ref{halo_extreme_sub1}). 

  The resulting SZE image in Fig.(\ref{halo_extreme_sub1}) clearly has many compact SZE
  features. This implies that the compact SZE feature seen in
  Fig.(\ref{nsrcmap2}) is not merely an artifact of simple (smooth) spatial spectral
  index variations. The figure clearly suggests that a different
  approach is needed in order to de--emphasize the compact SZE feature
  -- the radio halo must have a different variation in order for the
  compact feature to be absent. We explore the simplest such variation below.
\subsection{Constraints on Radio Halo properties needed to rule out a compact feature}
Since a simple de--emphasizing of the SZE in the previous section did
not yield an SZE image with no compact features, we now attempt to
find the radio halo distribution at 18 GHz, which, when subtracted
from our 18 GHz image, would yield a uniform SZE over the entire
region of the cluster. We do this analysis first along a line
connecting the SZM and the RHM, and then for the entire image, as
described below.
\begin{figure}
  \begin{center}
    \includegraphics[height=70mm, angle=0]{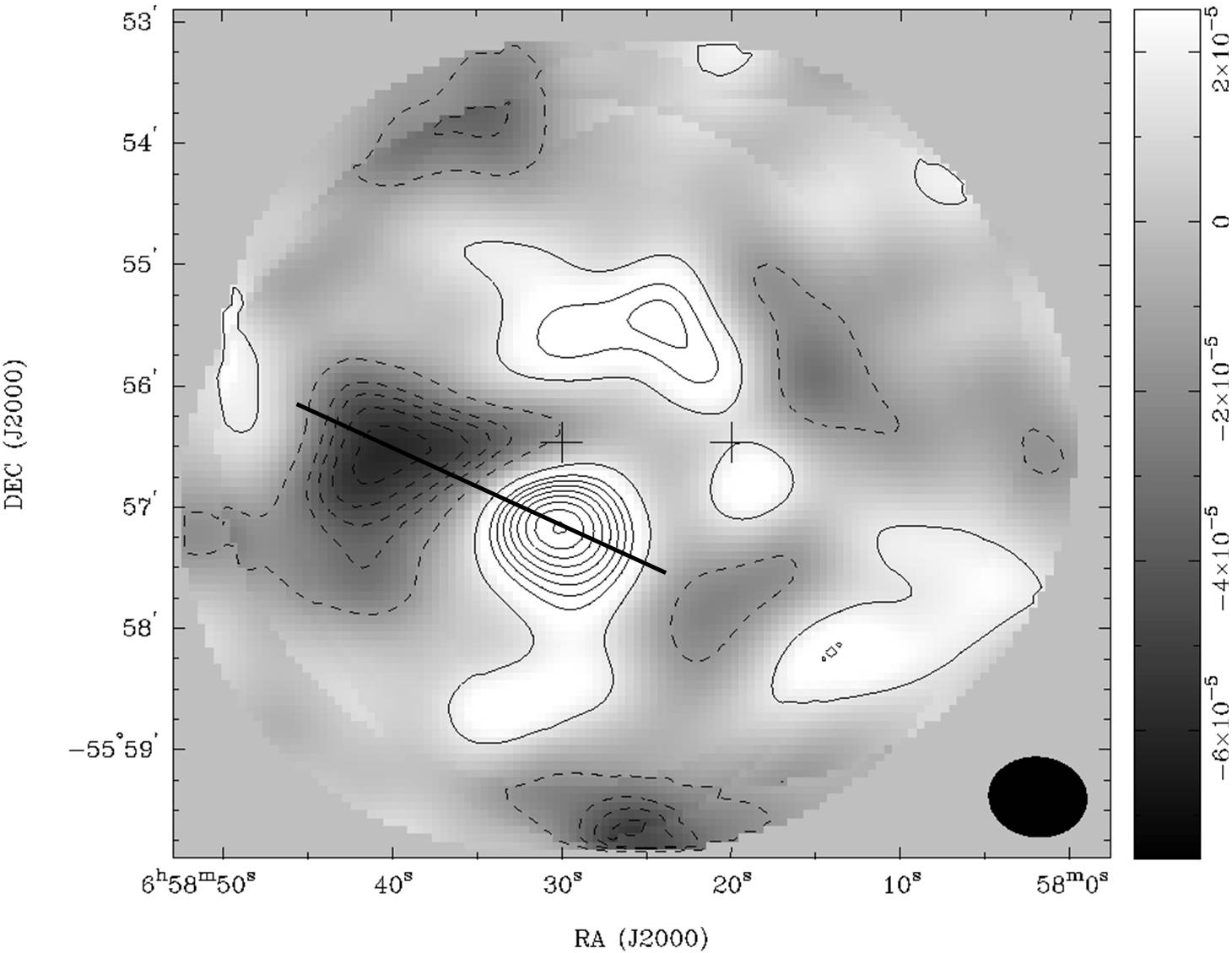}\\
    \caption{
      A slice connecting SZM and RHM, and including the entire extent
      of the SZE feature as well as the radio halo.
    }
    \label{45arcsec_slice}
  \end{center}
\end{figure}
\begin{figure}
  \begin{center}
    \includegraphics[height=70mm, angle=0]{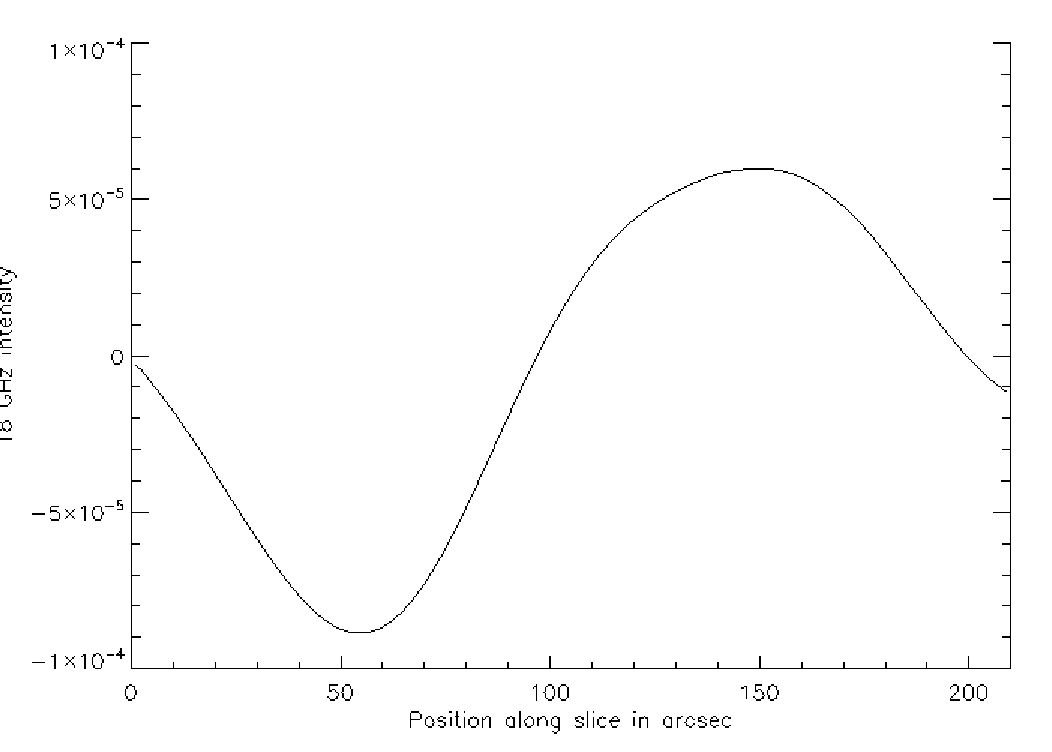}\\
    \caption{
      A plot of intensity along the slice shown in
      Fig.~\ref{45arcsec_slice}. This plot starts from the bottom
      right corner of the slice in Fig.~\ref{45arcsec_slice}.
    }
    \label{45arcsec_slice_1dplot}
  \end{center}
\end{figure}
\begin{figure}
  \begin{center}
    \includegraphics[height=70mm, angle=0]{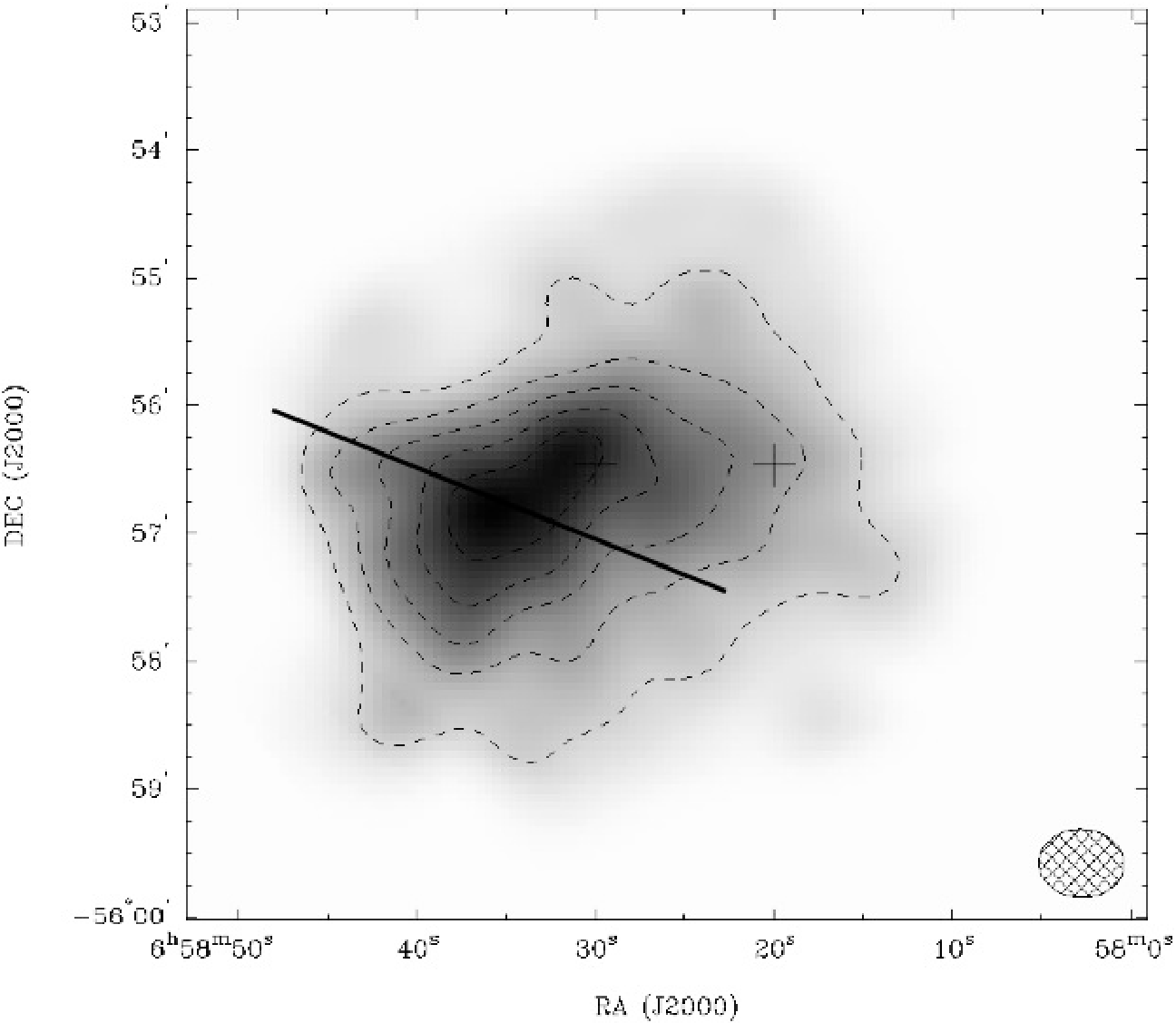}\\
    \caption{
      The same slice as in the previous figure, on the radio halo image.
    }
    \label{halo_spx_slice}
  \end{center}
\end{figure}
\begin{figure}
  \begin{center}
    \includegraphics[height=70mm, angle=0]{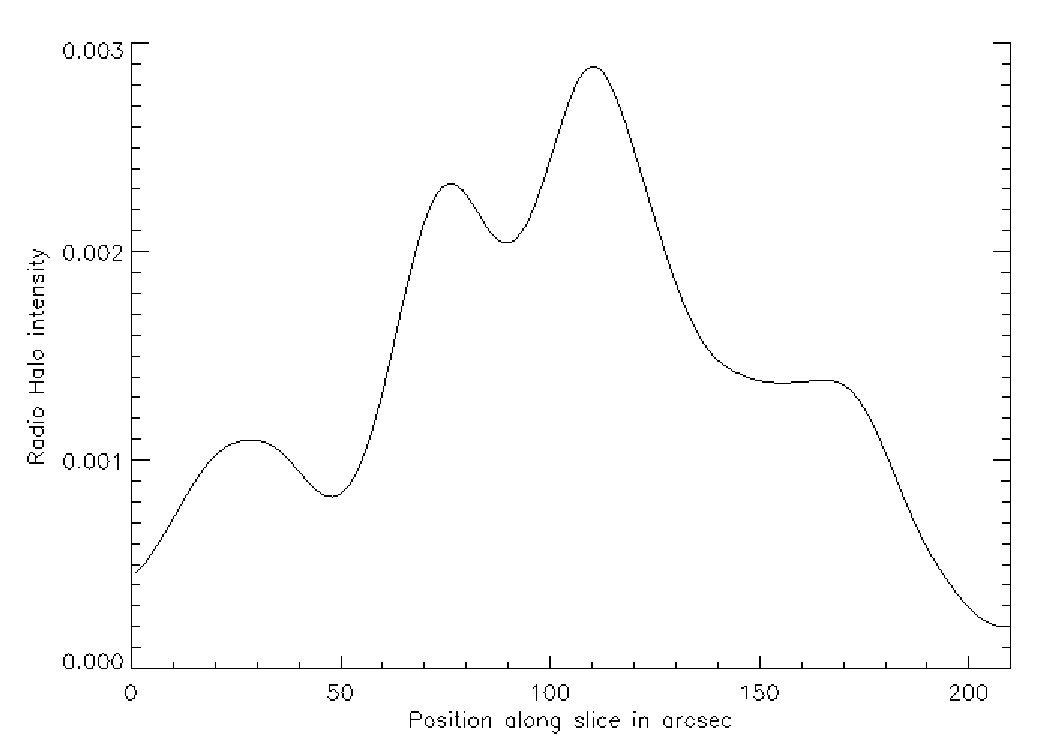}\\
    \caption{
      A plot of intensity along the slice shown in
      Fig.~\ref{halo_spx_slice}. This plot starts from the bottom
      right corner of the slice in Fig.~\ref{halo_spx_slice}.
    }
    \label{halo_spx_slice_1dplot}
  \end{center}
\end{figure}
\begin{figure}
  \begin{center}
    \includegraphics[height=70mm, angle=0]{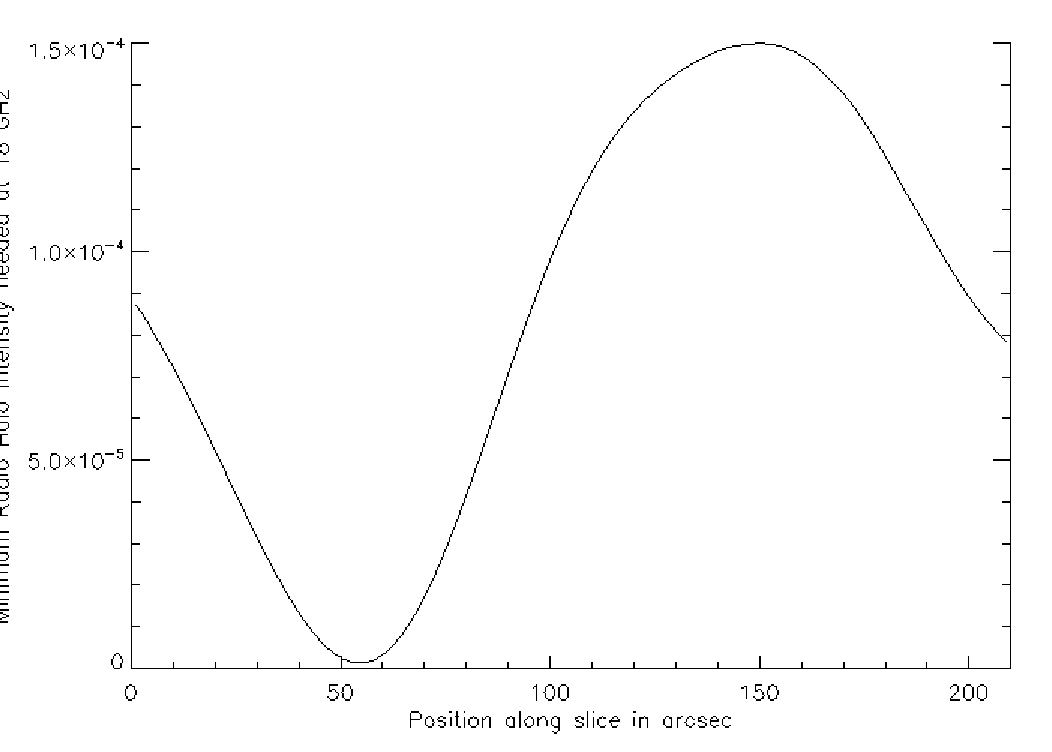}\\
    \caption{
      A plot of the minimum Radio Halo intensity needed to make the SZ
      in the 18 GHz image non--compact. This is obtained by
      subtracting the peak dip of the SZ feature from the radio halo
      intensity along the slice defined in the above figures.
    }
    \label{min_slice_1dplot}
  \end{center}
\end{figure}
\begin{figure}
  \begin{center}
    \includegraphics[height=60mm, angle=0]{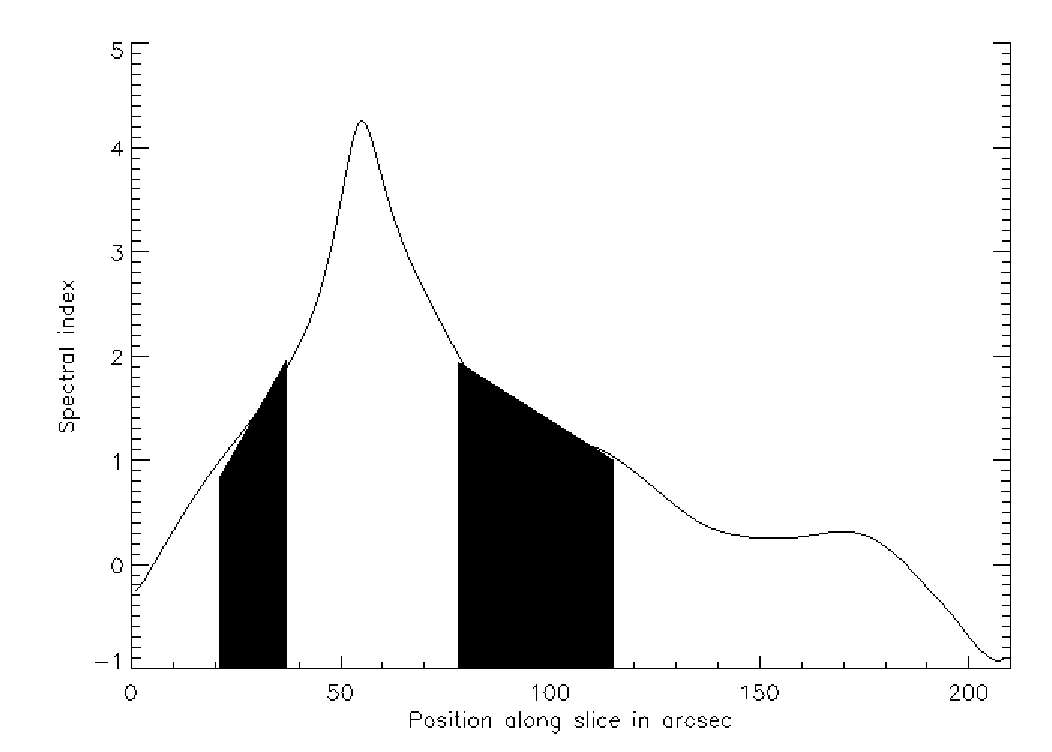}\\
    \caption{
      Variation of the spectral index (between 6 and 18 GHz) as a
      function of distance along slice which is needed to make the SZ
      feature in the 18 GHz image non--compact. This spectral index is
      derived from the minimum radio halo level needed to make the SZ
      feature non--compact, shown in Fig.~\ref{min_slice_1dplot}. The area that has $2>\alpha
      >1$ is marked in the figure. Notice that only $\sim$25\% of the
      plot has a spectral index $2>\alpha >1$. The region around SZM needs to have a very
      steep spectrum in order for the SZE feature to be
      non--compact. This kind of extreme variation in the spectral
      index ($4>\alpha >0.1$) has never been observed in any radio
      halo, because radio halos are produced by synchrotron. 
    }
    \label{index1}
  \end{center}
\end{figure}
\subsubsection{One--dimensional analysis}
\label{onedanalysis}
A slice connecting the SZM and the RHM was taken, as shown in
Fig.(\ref{45arcsec_slice}). A plot of intensity values along the slice
is shown in Fig.(\ref{45arcsec_slice_1dplot}). If the SZE was a large, diffuse feature,
it would be a constant over this slice. The {\em least} value of this
constant would be the same as the lowest value in the plot in
Fig.(\ref{45arcsec_slice_1dplot}), i.e. $-$90$\mu$Jy beam$^{-1}$. We therefore assume that the
diffuse SZE will have this value -- this will yield the {\em minimum}
spatial variations in the spectral index needed to have a non--compact
SZE. 

First, the above value ($-$90$\mu$Jy beam$^{-1}$) was subtracted from
the intensity vs. position plot, yielding the {\em minimum amount} of
radio halo emission at 18 GHz, as shown in Fig.(\ref{min_slice_1dplot}). Next,
the same slice was taken in the 1.344 GHz radio halo image, as shown
in Fig.(\ref{halo_spx_slice}). The spectral index may now be calculated
as a function of the minimal 18 GHz radio halo intensity along the slice, as shown in
Fig.(\ref{index1}) ($I_{18}$), 6 GHz intensity ($I_{6}$; since the spectral
index is known to be constant between 1.344 and 6 GHz, from Liang et al) and therefore, distance, from the following relations
\begin{equation}
I_{6}=I_{L}\left(\frac{6}{1.344}\right)^{-1.3}
\label{6l}
\end{equation}
\begin{equation}
I_{18}=I_{6}\left(\frac{18}{6}\right)^{-\alpha}=I_{L}\left(\frac{6}{1.344}\right)^{-1.3}\left(\frac{18}{6}\right)^{-\alpha}
\label{186}
\end{equation}
And so,
\begin{equation}
\alpha=\frac{\log(I_{18})-\log(I_{L}) + 1.3\log(\frac{6}{1.344})}{\log(\frac{1}{3})}
\label{spin1}
\end{equation}
$\alpha$ is plotted in Fig.(\ref{index1}). Most ($\sim$75\%) of the
points in the spectral index plot lie outside the known and observed
range $2>\alpha >1$. For a small subset of points in the plot, the
spectral index is inverted - this extreme/peculiar behaviour of the
spectral index is not possible to explain physically, through any
existing model or process. This extreme variation in the spectral
index is therefore not likely to occur, and indeed, is presented here
to emphasize the compact nature of the SZE in the Bullet cluster.
\begin{figure}
  \begin{center}
    \includegraphics[height=70mm, angle=0]{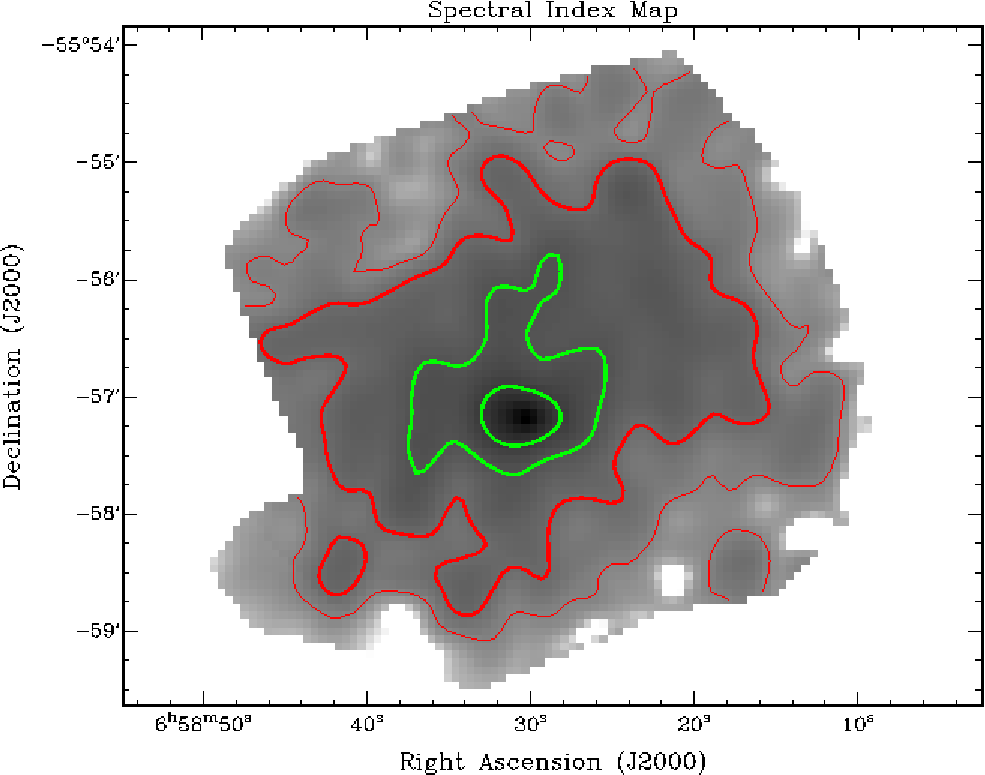}\\
    \caption{
      An image of the spatial variation of the spectral index (between 6 and 18 GHz) as a
      function of position. The area that has $2>\alpha
      >1$ is marked between the two green contours in the figure. Notice that only $\sim$25\% of the
      plot has a spectral index $2>\alpha >1$. The region around SZM needs to have a very
      steep spectrum in order for the SZE feature to be
      non--compact. This kind of extreme variation in the spectral
      index ($4>\alpha >0.1$) has never been observed in any radio
      halo, because radio halos are produced by synchrotron. 
    }
    \label{index_map1}
  \end{center}
\end{figure}
\begin{figure}
  \begin{center}
    \includegraphics[height=70mm, angle=0]{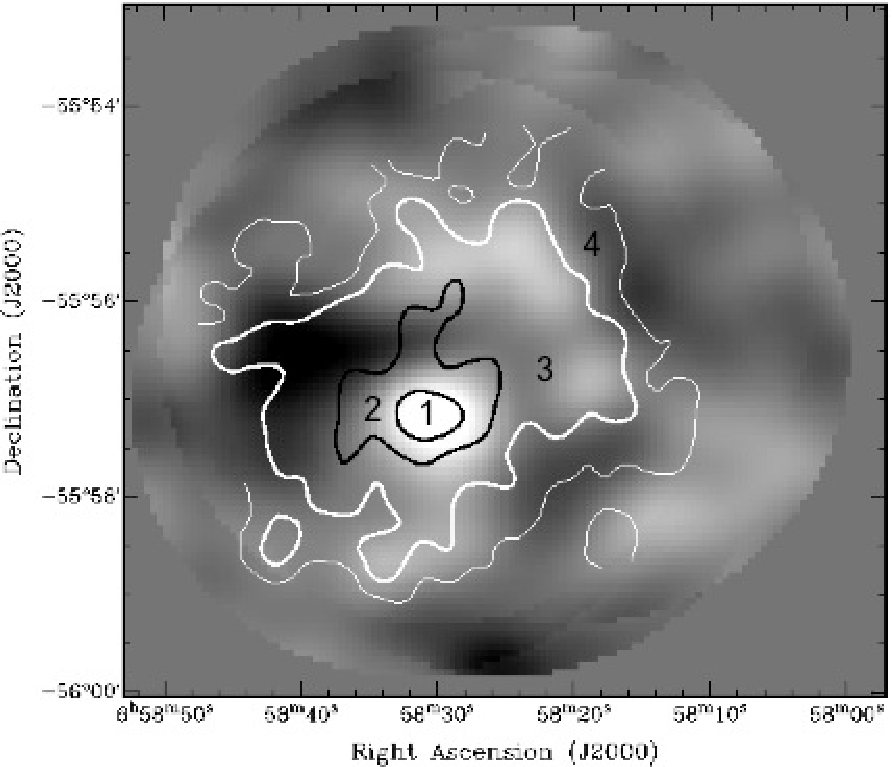}\\
    \caption{
      Our 18 GHz image, with an overlay of contours of the spectral
      index needed between 6 and 18 GHz in order to make the SZ
      feature non--compact. Regions with different ranges are marked
      in the figure as follows. Region 1: 4$\leq\alpha >$2; Region 2: 2$>\alpha >$1; Region 3: 1$>\alpha >$0;
Region 4: $\alpha >$0. Notice that region 2 has $<$25\% of the area in
    the index contour.}
    \label{index_map2}
  \end{center}
\end{figure}
\begin{figure}
  \begin{center}
    \includegraphics[height=70mm, angle=0]{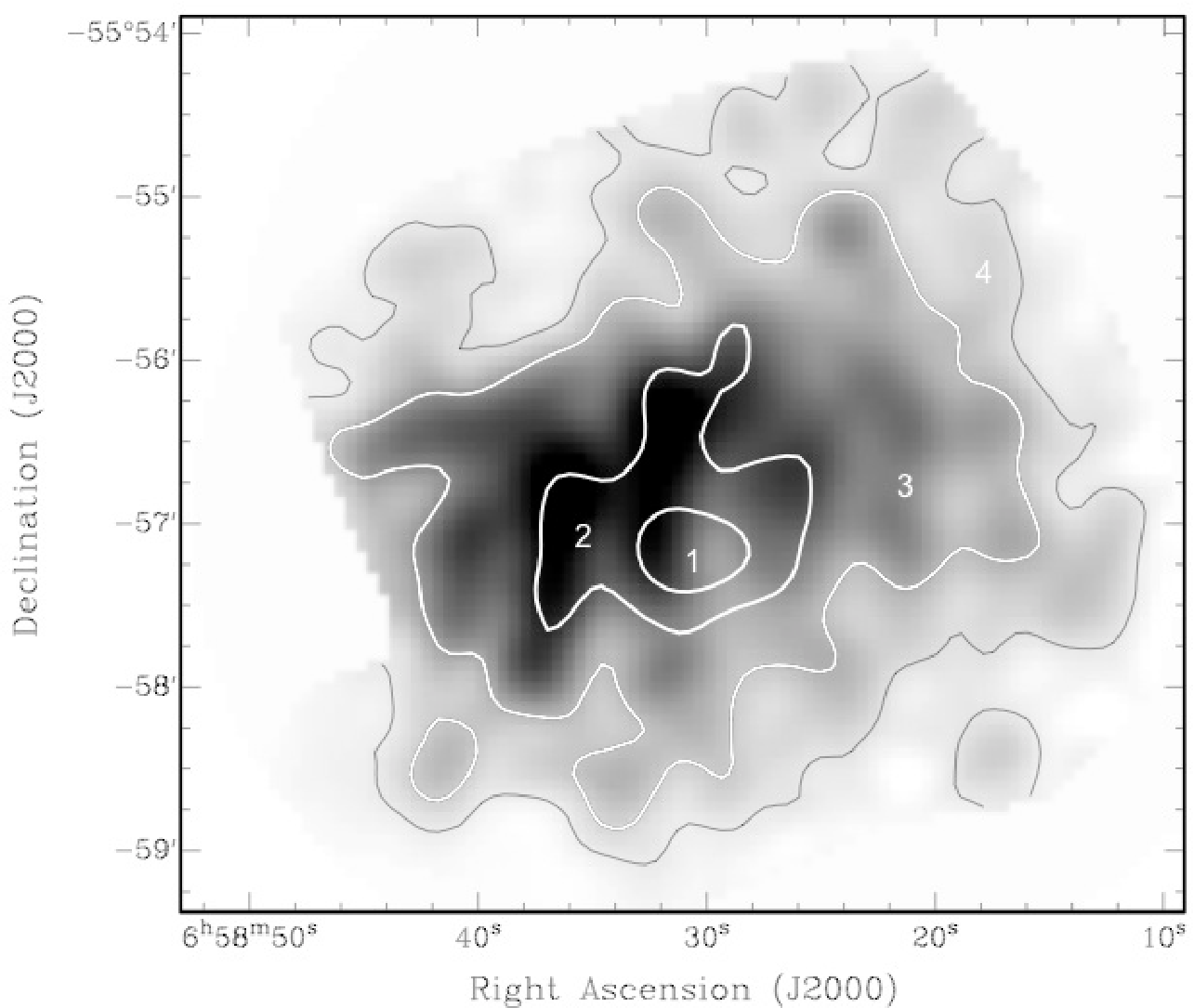}\\
    \caption{
      1.344 GHz image from \citep{2000ApJ...544..686L}, with an overlay of contours of the spectral
      index needed between 6 and 18 GHz in order to make the SZ
      feature non--compact. Regions with different ranges are marked
      in the figure as follows. Region 1: 4$\leq\alpha >$2; Region 2: 2$>\alpha >$1; Region 3: 1$>\alpha >$0;
Region 4: $\alpha >$0. Notice that region 2 has $<$25\% of the area in
    the index contour.}
    \label{index_map3}
  \end{center}
\end{figure}
\subsubsection{Image analysis}
\label{imageanalysis}
The process described in the previous section can be applied to the
entire 2--dimensional image instead of a single
slice. Eq.(\ref{spin1}) then denotes the spectral index $\alpha$ as a
function of (RA,Dec.). We applied eq.(\ref{spin1}) to
Fig.(\ref{nsrcmap2}), our 18 GHz image, and an extrapolation of
Fig.(\ref{halo1344spx}) to 6 GHz using the spectral index measured by
\citep{2000ApJ...544..686L}, in order to obtain the spectral index map
required for the SZ feature to be non--compact. This spectral index
map is shown in Fig.(\ref{index_map2}\&\ref{index_map3}). 

It is more useful to overplot the contours of this spatial variation of
the spectral index on our 18 GHz image as well as the radio halo;
these overplots are shown in Figs.(\ref{index_map2}\&\ref{index_map3})
respectively. From Fig.(\ref{index_map2}), it is clear that the region
with the deepest SZE coincides with the region that has the steepest
spectral index, 4.0$>\alpha >$2.0. This range for the spectral index
is far beyond the limits obtained from observations of radio halos
thus far. The reason that spectral indices this steep have never been
observed (and are not likely to be observed) is that radio halos are
caused by synchrotron emission. A steepening in radio halo emission
from the region with a deep SZE implies a steepening in the Energy
spectrum of electrons in that region. This is unlikely, given that
SZE$\propto\int n_{e}T_{e}d\ell$. Therefore, a spectral index
steepening in a deep SZE region is physically inconsistent. 

Additionally, spectral indices of $\alpha <$1 have not been observed,
whereas we find large regions with this spectral index in
Fig.(\ref{index_map2}). A relatively flat spectral index of 1$>\alpha
>$0.7 may be expected in the hottest and most turbulent regions in the
ICM of clusters that have undergone recent mergers. However,
flattening is not likely to occur a significant distance away from the
most turbulent and hottest regions of the ICM, which is what we
observe in the spectral index map in Fig.(\ref{index_map1}). 

We may therefore conclude that in trying to rule out a compact SZE
feature, we have admitted physically inconsistent ranges of spectral
indices in a large ($>$75\%) region of the ICM in the Bullet
cluster. This attempt to rule out a compact SZE feature is also,
therefore, a failure.

In order to obtain further constraints on the SZE and the radio halo,
the two components need to be separated through multi--frequency
observations; a comparison of 18 GHz and 150 GHz images can lead to
SZE--halo separation; however, uv--coverages of the ATCA and APEX
instruments is complementary (since APEX is a single--dish
experiment). Single--dish measurements of the Bullet cluster at 18 GHz
are therefore needed to separate radio halo and SZE.
\begin{figure}
  \begin{center}
    \includegraphics[height=70mm, angle=0]{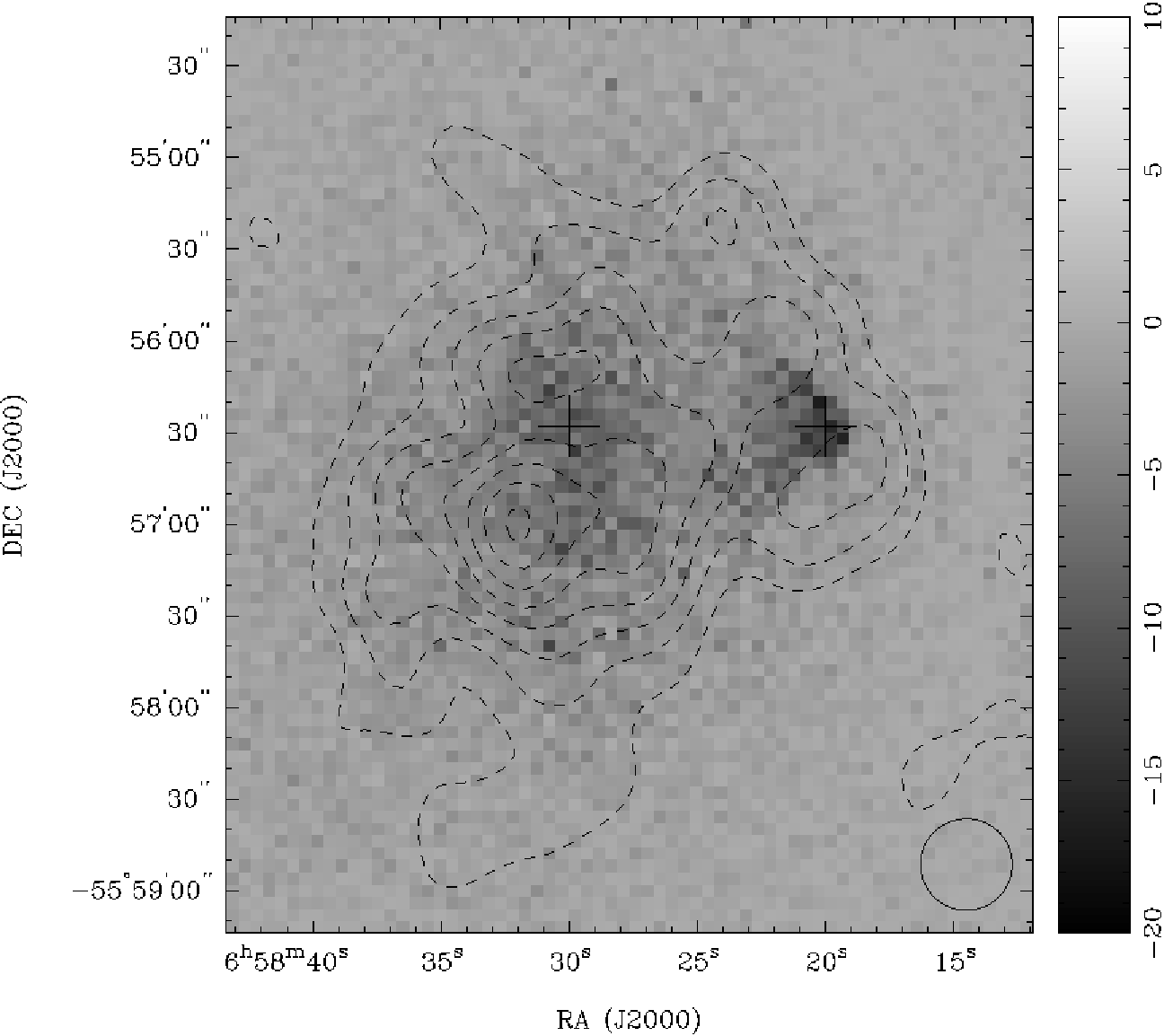}\\
    \caption{
      X--ray greyscale overlaid with contours of SZE obtained after
      subtracting the radio halo assuming a spectral index of
      $\alpha=$1.3 from 1.344 to 18 GHz. Contour levels are at
      ($-$2,$-$4,$-$6,$-$8,$-$10,$-$12,$-$14,$-$16) times the noise
      RMS of 6.5$\mu$Jy beam$^{-1}$.
    }
    \label{nohxray1}
  \end{center}
\end{figure}

\section{A discussion of the SZE in the Bullet cluster}
\label{sz1}
The sky distribution in the SZE does indeed coincide with the radio halo
distribution in the western and southen parts; additionally, in our
high-angular resolution interferometer image, SZE is also detected
towards the western end of the radio halo in two distinct features in
the NW and SW parts of the image. The significance of the peak dips in
these features is 4.0$\sigma$ and 5.5$\sigma$ (times the noise RMS of
8.0$\mu$Jy beam$^{-1}$) respectively, in Fig.(\ref{nsrcmap2}) --
without radio halo subtraction. 

We emphasize that the SZE appears as a compact feature. The deepest SZE feature is at 
RA: $06^{\rm h}58^{\rm m}30\fs285$, DEC: $-55\degr57\arcmin12\arcsec$ in
Fig.(\ref{nsrcmap2}) and is clearly offset $65\arcsec$ to the SE of the eastern X-ray emission peak; this 
deep SZE hole is a flux density decrement of $-$88.8$\mu$Jy/beam

Significantly differing depths in the SZE
decrement in the APEX and ATCA images imply that there is substantial extended
SZE structure in the Bullet cluster on angular scales of $2\farcm6$ and
larger, which is resolved out and missing in the ATCA inteferometer image.  

Clearly, the data suggest the existence of a range of scales in the
SZE structure in this merging cluster, requiring imaging with
sensitivity to a range of spatial scales to produce a holistic image
of the SZE. In Fig.~\ref{nohxray1} we show contours of the SZE overlaid on Chandra X-ray
intensity distribution shown using grey scales.

We discuss in detail below the SZE features in our 18 GHz image:
\begin{enumerate}
\item We observe a deep SZE `hole' significantly displaced
  to the SE of the X-ray peak in the hotter eastern cluster. Since SZE
  probes integrated pressure along the line of sight, this
  suggests that there are significant pressure variations in the
  intracluster gas, presumably because of the ongoing cluster collision. 
  X--ray spectral luminosity is proportional to $\int n_{\rm{e}}^{2} 
  T^{-\frac{1}{2}}d\rm{l}$, whereas the SZE is proportional to $\int
  n_{\rm{e}}T d\rm{l}$: the two effects represent different weightings of
  temperature $T$ and electron density $n_{\rm{e}}$. X-ray emission images
 are more representative of $n_{\rm{e}}$ distribution, whereas the SZE images
 are representative of $T$ distribution. 
  The offset of the SZE `hole' from the emission peak suggests that the peak in intracluster gas
  temperature is spatially offset from the peak in gas density in the Bullet
  cluster\footnote{\citet{2002ApJ...567L..27M} have indeed reported finding highest temperatures
  SE of the X--ray peak.}. 
  RXJ1347$-$1145 \citep{2001PASJ...53...57K,2010ApJ...716..739M} and
  CL0152$-$1357 \citep{2010ApJ...718L..23M} are the only other clusters that display this offset, which is a demonstration of the complexity in intracluster gas distribution properties
  in merging clusters.
  We also observe significant SZE substructure
  across the cluster complex. All these SZE components are displaced from the
  peaks in X-ray emission corresponding to the two clusters comprising the
  Bullet cluster complex. 
\item All substructure we detect in the SZE image lies within the boundary of the
  radio halo. Moreover, careful examination of Fig.(\ref{nohc30}) indicates
  that the SZE in the western parts of the cluster complex appear to follow
  the contours of the radio halo. This has also been seen by \citet{2011A&A...527L...1C}. This is surprising, since the radio halo arises from synchrotron plasma and the SZE from thermal intracluster gas and
  there is no spatial correspondence expected between these distinct
  components. A plausible reason for the cospatial locations of the SZE and
  radio halo---in the western parts of the cluster---is a common origin for
  the two populations of radiative electrons.  Radio halos are believed to be
  synchrotron emission from electrons accelerated (or re-accelerated) to
 relativistic energies in the turbulence generated by cluster mergers.  It
  may be that the radio halo was created in the turbulent wake of the passage
  of the `Bullet' through the cluster gas, and the SZE also arises cospatially
  from relatively hotter electrons in the same turbulent wake. Absence of SZE
  towards the eastern parts of the radio halo would, in this picture, indicate
  longer lifetimes for the relativistic electrons compared to the SZE
  electrons. \citep{2002ApJ...567L..27M} have reported finding spatial correlation between halo
  brightness and local gas temperature in merging clusters; this is
  supplementary to the SZE and the halo being cospatial in our observations.
\item SZE dip is relatively shallow towards the peaks in the X-ray
  emission.  In Fig.~\ref{nohxray1} it may be noted that the SZE contours are
  indented in sky regions where the X-ray emission is most intense. The observed avoidance of the highest emission
  measure regions by SZE suggests that the highest density parts of the
  cluster gas may have somewhat cooler temperatures, owing to shorter cooling
  timescales, and consequently contribute relatively less to the SZE.
\item The NW and SW SZE features are close to the shock fronts in the
  X--ray image, as is clear from Fig.~\ref{haloxray2}. However, the
  SZE feature dips appear in front of the shock fronts, not behind --
  but shock fronts are the sites with highest pressure, and pressure
  peaks cannot appear in front of shocks. We may infer from this that
  the pressure peaks are behind or at the shock fronts, but due to
  no radio halo subtraction, the SZE dips appear to be in
  front of the shocks. 
\end{enumerate}
\citet{2010ApJ...718L..23M} have recently found a similar displacement of SZE dip from X--ray brightness peaks in the merging
cluster CL0152$-$1357. In addition, this effect has been known to
exist in the merging cluster RXJ1347$-$1145 \citep{2001PASJ...53...57K,2010ApJ...716..739M}.

\section{Summary}
\label{summary}
We have presented 18~GHz observations of the Bullet cluster from the
ATCA, and report detection of substructure in the Sunyaev$-$Zeldovich
Effect. We rn through several tests of consistency to prove that the
image of the SZE in the Bullet cluster is robust. We subtract unresolved continuum sources in the cluster field using
18~GHz ATCA images made with enhanced weightings to the longer visibility
spacings. We run through two ways of ruling out a compact feature in
the SZE, and arrive at the conclusion that inconsistent and non--physical steepening
and flattening of the radio halo spectral index must be resorted to in
order to rule out a compact SZE feature. We show that the most prominent SZ effect feature in the 18
GHz image must be a compact feature, given reasonable assumptions
about the Radio Halo. This deep SZE feature is significantly displaced from the peak in X-ray emission.
The main conclusions from this work are:
\begin{enumerate} 
\item High--frequency ($\nu >$10 GHz) observations are crucial for
  probing the dynamics of energtic cluster mergers. 
\item The intracluster gas in the merging
Bullet cluster appears to have significant pressure distribution structure
that differs from that in gas emission measure, galaxy and dark matter
distributions.
\end{enumerate}
Future ALMA observations with a range of spatial frequencies would be
expected to provide SZE images that reproduce the compact and extended
structure together and hence be amenable to modeling of the gas in
such merging clusters, where the pressure and temperature may have
complex distribution structures. However, it may be noted that the
synthesized of even the most compact sub--array in ALMA is too
small, even at the lowest frequencies at which ALMA operates, to
characterize the structures in the Bullet cluster (e.g. ALMA
field--of--view at 100 GHz, the lowest operating frequency, is only
60$\arcsec$; the smallest baselines of $\sim$18 m yield an angular
scale of 35$\arcsec$). The total--power mode of ALMA is the only way
that the large--scale structure of the SZE in the Bullet cluster can
be characterized.

In order to enable a proper difference between the large--scale SZE at 150
GHz from APEX and the SZE at 18 GHz, leading to a characterization of
the radio halo, a single--dish observation of the
Bullet cluster at 18 GHz is needed. This measurement is possible using
the 64 m Parkes Telescope in Australia. 
\section*{Acknowledgments}
The Australia Telescope
Compact Array is part of the Australia Telescope which is funded by
the Commonwealth of Australia for operation as a National Facility
managed by CSIRO. 

Greyscale in Figure~\ref{nohxray1} courtesy of the Chandra X--ray Observatory
Center, which is operated by the Smithsonian Astrophysical Observatory
on behalf of NASA.

Travel to Australia was made possible by two travel grants from the
Raman Research Institute, Bangalore, India and one from IUCAA, Pune, India.

We gratefully acknowledge a part of the observations being carried out by Dr. R. Subrahmanyan and Dr. M. Wieringa. We also thank Dr. D. Narasimha for motivating the search for SZ polarization. 

SSM gratefully acknowledges the support provided by CSIR (Council for Scientific and Industrial Research), Government of India, through their grant number 03(1462)19\_EMR$-$II.

\bibliographystyle{mn2e}

\end{document}